\documentclass[12pt]{article}
\usepackage[xdvi]{graphicx}
\newcommand{\beq}{\begin{equation}}
\newcommand{\eeq}{\end{equation}}
\newcommand{\bea}{\begin{eqnarray}}
\newcommand{\eea}{\end{eqnarray}}

\def\e2sig{e^{-2r\sigma}}

\newcommand{\ccr}{c_{\textrm{\scriptsize cr}}}

\begin{document}

\begin{titlepage}
%%%%% PREPRINT NUMBERS %%%%%%
\begin{flushright}
ROME1/1423/06 \\
TIT/HEP-549 \\
hep-th/0602123
\end{flushright}
\vspace*{2cm}
%%%%%%%%%%%%%%%%%%% TITLE %%%%%%%%%%%%%%%%%%
\begin{center}{\Large\bf Supersymmetry Breaking \\
by Constant Boundary Superpotentials \\
in Warped Space}
\end{center}
%%%%%%%%%%%%%%%% AUTHORS %%%%%%%%%%%%%%%%%%%%%%%
\vspace{1cm}
\begin{center}
{\bf Nobuhito Maru}$^{(a)}$
\footnote{E-mail: Nobuhito.Maru@roma1.infn.it},
{\bf Norisuke Sakai}$^{(b)}$
\footnote{E-mail: nsakai@th.phys.titech.ac.jp},
~and~~{\bf Nobuhiro Uekusa}$^{(b)}$
\footnote{E-mail: uekusa@th.phys.titech.ac.jp}
\end{center}
%%%%%%%%%%%%%%%%%%%%%%% AFFILIATION %%%%%%%%%%%%
\vspace{0.2cm}
\begin{center}
%\small
${}^{(a)}$ {\it Dipartimento di Fisica, 
Universit\`a di Roma "La Sapienza" \\and INFN, Sezione di Roma, 
P.le Aldo Moro 2, I-00185 Roma, Italy}
\\[0.2cm]
${}^{(b)}$ {\it Department of Physics, Tokyo Institute of Technology, 
Tokyo 152-8551, Japan}
%%%%%
%%%%%%%
\end{center}
%%%%%%%%%%%%%%%%%% ABSTRACT %%%%%%%%%%%%%%%
\vspace{1cm}
\begin{abstract}
Supersymmetry breaking by constant (field independent) 
superpotentials localized at boundaries is studied 
in a supersymmetric warped space %Randall-Sundrum 
model. 
We calculate the Kaluza-Klein mass spectrum of the 
hypermultiplet. 
We take into account of the radion and the compensator 
supermultiplets, as well as the bulk mass $c$ for the 
hypermultiplet. 
The mass splitting is similar to that of the 
Scherk-Schwarz supersymmetry breaking (in flat space) 
for large $|c|$, and has an interesting dependence on 
the bulk mass parameter $c$. 
We show that the radius is stabilized by the presence of 
the constant boundary superpotentials. 
%The possibility of the radius 
%stabilization is also discussed. 

\end{abstract}
\end{titlepage}
%%%%%%%%%%
%\tableofcontents
%%%%%%%%%%%%
\newpage
\renewcommand{\theequation}{\thesection.\arabic{equation}} 
%%%%%%%%%%%%%%%%%%%%%%%%%%%%%%%%%%%%%%%%%%%%%%%%%%%%%%%%%%
\setcounter{equation}{0}
%%%%%%%%%%%%%%%%%%%%%%
\section{Introduction}
%%%%%%%%%%%%%%%%%%%%%%
Supersymmetry (SUSY) \cite{SUSY} is a well-motivated 
extension to the Standard Model, which plays a crucial 
role in solving the gauge hierarchy problem. 
Extra dimensions with flat space \cite{ADD}, 
\cite{Cremades:2002dh} or with the 
warped space \cite{RS} are also an alternative solution 
to the gauge hierarchy problem. 
Considering both ingredients is natural in the context 
of the string theory and is often taken as the starting 
point in the phenomenological model of the brane world 
scenarios. 
In such a setup, we have to compactify extra dimensions 
and break SUSY to obtain realistic four-dimensional 
physics. 
One of the simple ways to realize it is the Scherk-Schwarz 
(SS) mechanism of SUSY breaking \cite{SS}.

It is known that the SS SUSY breaking is equivalent 
to the SUSY breaking by a (bulk) constant (field independent) superpotential 
in flat space \cite{MP, BFZ, GR}. 
These two scenarios generate the same mass spectrum. 
It is natural to ask whether this equivalence still holds 
in warped space. This issue has been discussed in several 
interesting papers so far \cite{HNOO, BB, AS}. 
According to \cite{HNOO}, if one considers the gauging of 
a symmetry of the theory in warped space, 
which is completely broken by the boundary conditions, 
one is led to an inconsistent theory. 
This implies an inconsistency of the SS SUSY breaking in a 
SUSY %warped space %
Randall-Sundrum 
model. 
In \cite{BB}, they discuss the SS twist for $SU(2)_R$ 
in the five-dimensional gauged supergravity. 
Their conclusion is that if the background geometry is 
$AdS_4$, SUSY is broken by the SS twist, but if the 
background geometry is the %warped %
Randall-Sundrum 
geometry, 
SUSY is not broken by the SS twist. 
This statement agrees with that of \cite{HNOO}. 
Recently, \cite{AS} also discussed this issue from the 
viewpoint of the five-dimensional conformal supergravity. 
It tells us that whether SUSY is broken by the SS twist 
or not in warped space depends on the way to gauge $U(1)_R$. 
If we gauge $U(1)_R$ by the graviphoton with 
$Z_2$-odd gauge coupling, SUSY is not broken by the SS twist. 
However, they pointed out that if we gauge $U(1)_R$ by 
the graviphoton with $Z_2$-even gauge coupling, it is 
possible to break SUSY. 
This seems to disagree with the statement in 
\cite{HNOO,BB}. 
In the light of these facts, the issue of the SS SUSY 
breaking in warped space is not settled. 
Furthermore, the equivalence of the SS SUSY breaking 
to the SUSY breaking by a constant (field independent) 
superpotential 
in warped space is also unclear compared to the flat 
space case. 
A number of works have studied the SUSY %warped space 
Randall-Sundrum 
model \cite{ABN, GP, FLP, BKV}. 

More recently, higher dimensional gauge theories are 
also used to provide Higgs fields as an extra-dimensional 
component of gauge fields \cite{Hatanaka:1998yp}. 
The true ground state of such a system is generally 
determined by the one-loop effective potential
\cite{Haba:2003ux}--\cite{Hosotani:2006qp}, which 
requires mass spectrum of all the Kaluza-Klein towers, 
especially their SUSY breaking pattern, since the 
effective potential vanishes in the supersymmetric 
limit. 
For phenomenological applications, it is also important 
to include bulk mass parameters for 
hypermultiplets \cite{Hosotani:2006qp}.

The purpose of this paper is to investigate SUSY 
breaking effects and its properties caused by constant 
(field independent) superpotentials localized 
at fixed points in the SUSY %warped space %
Randall-Sundrum 
model. 
Taking the hypermultiplet and including the compensating 
multiplet and the radion multiplet consistently 
(from supergravity), we calculate the Kaluza-Klein mass 
spectrum and their mode functions of the hypermultiplet. 
We find that the mass spectrum depends on the bulk mass 
parameter in addition to the strength of the 
constant boundary superpotential. 
We observe a similarity of our result to the 
SS SUSY breaking spectrum in flat space for large bulk 
mass parameter $|c|$. 
%We also discuss the possibility of the radius stabilization 
We also show that the radius is stabilized by the presence of 
the constant boundary superpotentials 
\footnote{For related discussions on stabilization, 
see \cite{MO, MO1, EMS}, for example.}.

This paper is organized as follows. 
The model is introduced in Sec.\ref{sc:model}. 
The mass spectrum of the 
hypermultiplet is calculated in Sec.\ref{sc:mass} and 
is compared with the SS SUSY breaking. 
The mode functions are obtained in 
Sec.\ref{sc:mode_function}. 
We show in 
Sec.\ref{sc:radius_st} 
that the radius is stabilized. 
%the possibility of the radius stabilization. 
Sec.\ref{sc:conclusion} gives a conclusion. 
Appendices include a few details of calculations. 

\setcounter{equation}{0}
%%%%%%%%%%%%%%%
\section{Model}\label{sc:model}
%%%%%%%%%%%%%%%
We consider a five-dimensional supersymmetric model of 
a single hypermultiplet on the %warped space %
Randall-Sundrum 
background, 
whose metric is 
\bea
ds^2 = e^{-2R\sigma}\eta_{\mu\nu}dx^\mu dx^\nu +R^2 dy^2, 
\quad 
\sigma(y)\equiv k|y|, 
\eea
where $R$ is the radius of $S^1$ of the orbifold $S^1/Z_2$, 
$k$ is the $AdS_5$ curvature scale, and the angle of $S^1$ 
is denoted by $y(0 \le y \le \pi)$. 
In terms of superfields for four manifest supersymmetry, 
our Lagrangian reads \cite{MP}
\bea
{\cal L}_5 &=& \int d^4 \theta 
\frac{1}{2} \varphi^\dag \varphi (T+T^\dag) 
e^{-(T+T^\dag)\sigma}
(\Phi^\dag \Phi + \Phi^c \Phi^{c\dag} - 6M_5^3) 
\nonumber \\
&& + \int d^2 \theta 
\left[
\varphi^3 e^{-3T \sigma} \left\{
\Phi^c \left[
\partial_y - \left( \frac{3}{2} - c \right)T \sigma' 
\right] \Phi + W_b
\right\} + {\rm h.c.}
\right] ,
\label{lagrangian}
\eea
where $\varphi = 1 + \theta^2 F_{\varphi}$ is the 
compensator chiral supermultiplet (of supergravity), 
$T=R + \theta^2 F_T$ is the radion chiral supermultiplet, 
and $\Phi, \Phi^c$ are chiral supermultiplets representing 
the hypermultiplet. 
The $Z_2$ parity is assigned to be even (odd) for 
$\Phi (\Phi^c)$. 
The derivative with respect to $y$ is denoted by $'$, 
such as $\sigma'\equiv d\sigma/dy$. 
The five-dimensional Planck mass is denoted as $M_5$. 
Here we consider a model with constant (field independent) 
superpotentials localized at the fixed points $y=0, \pi$ 
\bea
W_b \equiv 2M_5^3 (w_0 \delta(y) + w_\pi \delta(y-\pi)), 
\label{eq:boundary_pot}
\eea
where $w_{0},w_{\pi}$ are dimensionless constants which 
will be assumed to be ${\cal O}(1)$.

Since interesting physics in the extra dimensions is 
contained solely in the part of the Lagrangian 
(\ref{lagrangian}) containing auxiliary components 
$F, F^c, F_T, F_\varphi$,
we extract that part 
\bea
{\cal L}_{{\rm aux}} &=& 
\left(
\frac{1}{2}e^{-2R\sigma}(2RF^\dag F + F_T F^\dag \phi 
+ F_T^\dag F \phi^\dag)
\right. \nonumber \\
&& \left. + \left\{
\frac{1}{2}e^{-2R\sigma}(2R \phi^\dag F %\sigma 
+ F_T(\phi^\dag \phi -3M_5^3))
(F_{\varphi}^\dag - F_T^\dag \sigma)
+ {\rm h.c.} \right\} 
+(\phi \leftrightarrow \phi^c) \right) \nonumber \\
&&+e^{-2R\sigma}R(\phi^\dag \phi + \phi^c \phi^{c\dag} - 6M_5^3)
(F_{\varphi}^\dag - F_T^\dag \sigma)
(F_{\varphi} - F_T \sigma) \nonumber \\
&& + \left[
3e^{-3R\sigma}(F_{\varphi} - F_T \sigma)
\left\{
\phi^c \left[ \partial_y 
-\left( \frac{3}{2} - c \right) R \sigma' \right]
\phi + W_b 
\right\} \right. \nonumber \\
&&\left. +e^{-3R\sigma}\left\{
F^c \left[ \partial_y 
-\left( \frac{3}{2} - c \right) R \sigma' \right]\phi
+ \phi^c \left[ \partial_y 
-\left( \frac{3}{2} - c \right) R \sigma' \right]F 
\right. \right. \nonumber \\
&& \left. \left.  
-\phi^c \left( \frac{3}{2} - c \right) F_T \sigma' \phi 
\right\} +{\rm h.c.}
\right],
\label{aux}
\eea
where $\phi^c$ and $F^c$ ($\phi$ and $F$) are scalar 
and auxiliary components of $\Phi^c$ ($\Phi$). 
The Lagrangian (\ref{aux}) gives the following equations of 
motion for auxiliary fields 
\bea
F &\!\!\!=&\!\!\! -\frac{e^{-R\sigma}}{R}
\left[
-\partial_y \phi^{c\dag} 
+ \left( \frac{3}{2} + c \right)R \sigma' \phi^{c\dag}
+\frac{\phi}{2M_5^3}W_b \right. \nonumber \\
&\!\!\!&\!\!\! \left. +\frac{1}{6M_5^3}\phi^\dag \phi 
\partial_y \phi^{c\dag} 
+\frac{1}{3M_5^3}\phi^{c\dag} \phi 
\partial_y \phi^{\dag} 
-\frac{1}{6M_5^3}\phi^\dag \phi 
\phi^{c\dag}\left( \frac{9}{2} -c \right)
R \sigma' 
\right], \label{Feom} \\
F^c &\!\!\!=&\!\!\! -\frac{e^{-R\sigma}}{R}
\left[
\partial_y \phi^{\dag} 
- \left( \frac{3}{2} - c \right)R \sigma' \phi^{\dag}
+\frac{\phi^c}{2M_5^3}W_b \right. \nonumber \\
&\!\!\!&\!\!\! \left. +\frac{1}{6M_5^3}\phi^c \phi^\dag 
\partial_y \phi^{c\dag} 
+\frac{1}{3M_5^3}\phi^{c\dag} \phi^c 
\partial_y \phi^{\dag} 
-\frac{1}{6M_5^3}\phi^c \phi^\dag 
\phi^{c\dag} \left( \frac{9}{2} -c \right)
R \sigma' 
\right], \label{Fceom} \\
F_{\varphi} &\!\!\!=&\!\!\! -\frac{e^{-R\sigma}}{R}
\left[
-\frac{1}{6M_5^3}\phi^\dag \partial_y \phi^{c\dag} 
-\frac{1}{3M_5^3}\phi^{c\dag}\partial_y \phi^\dag 
+\frac{1}{6M_5^3} \phi^\dag \phi^{c\dag}
\left( \frac{9}{2} -c \right)R\sigma'
-\frac{1}{2M_5^3}W_b 
\right. \nonumber \\
&\!\!\!&\!\!\! \left.
-\frac{3(1-2R\sigma)}{r}\phi^{c\dag} 
\partial_y \phi^\dag -\frac{3(1-2R\sigma)}{r}W_b 
+\frac{1-2R\sigma}{r}
\phi^{c\dag}\phi^\dag \left(\frac{3}{2} -c \right)R\sigma'
\right], \label{Fpeom} \nonumber \\
\\
F_T &\!\!\!=&\!\!\! -\frac{e^{-R\sigma}}{r}
\left[
6\phi^{c\dag} \partial_y \phi^\dag -2\phi^{c\dag}\phi^\dag 
\left(\frac{3}{2} -c \right)R\sigma' +6W_b
\right], \label{FTeom}
\eea
where the partial integration has been performed in 
(\ref{Feom}) and 
$r \equiv \phi^\dag \phi + \phi^{c\dag} \phi^c - 6M_5^3$. 
We can eliminate these auxiliary fields in (\ref{aux}) by 
substituting (\ref{Feom})--(\ref{FTeom}). 
To obtain the mass spectrum, we take out only the part 
of the Lagrangian which is bilinear in scalar fields 
\bea
&&{\cal L}_{{\rm aux}}^{\rm bilinear}
= - \frac{1}{R} e^{-4R\sigma} 
\left[
\left| -\partial_y \phi^{c\dag} 
+ \left( \frac{3}{2} + c \right) 
R \sigma' \phi^{c \dag} +\frac{\phi}{2M^3} W_b \right|^2 
\right. \nonumber \\
&& \left. + \left| \partial_y \phi^\dag 
-\left( \frac{3}{2} -c \right) 
R \sigma' \phi^\dag +\frac{\phi^c}{2M^3}W_b \right|^2
+ \left\{ \frac{W_b}{M^3}
\left( \frac{3}{2} -c \right) R \sigma' \phi \phi^c 
+ {\rm h.c.} \right\}
\right] . 
\label{auxap}
\eea
The bilinear part of the Lagrangian turns out to be identical 
to that derived from another Lagrangian 
\begin{eqnarray}
 {\cal L}_5 &=& \int d^4\theta {1\over 2} 
    \varphi^{\dagger} \varphi (T+T^{\dagger})
   e^{-2R\sigma}
    (\Phi^{\dagger}\Phi +\Phi^c\Phi^{c\dagger} -6M_5^3)
\nonumber
\\
   &+& \int d^2\theta \left[
    \varphi^{2} e^{-3R\sigma}
   \left\{\Phi^c\left[\partial_y
-\left({3\over 2}-c\right)T\sigma'\right] \Phi
    +W\right\} +\textrm{h.c.}\right] , 
\end{eqnarray}
which was proposed in Ref.\cite{AS1} based on a more accurate 
treatment of the radion superfield using supergravity.

\setcounter{equation}{0}
%%%%%%%%%%%%%%%%%%%%%%%%%%%%%%%
\section{Mass spectrum of hyperscalar}\label{sc:mass}
%%%%%%%%%%%%%%%%%%%%%%%%%%%%%%%
%\subsection{Eigenvalue equations}\label{sc:eigenvalue}

Let us calculate the mass spectrum of scalar component 
 fields $\phi$ and $\phi^c$ of the hypermultiplet. 
To allow possible discontinuities of the $Z_2$ odd field 
$\phi^c$ across the fixed points $y=0, \pi$, we define 
\begin{equation}
\phi^c(x,y) \equiv \hat \epsilon(y) h^c(x,y), 
\qquad 
\hat \epsilon(y)\equiv 
\left\{
\begin{array}{cc}
+1, & 0<y<\pi \\
-1, & -\pi<y<0
\end{array}
\right.
, 
\label{eq:odd_field}
\end{equation}
where $h^c(x,y)$ is a parity even function with possibly 
nonvanishing value at $y=0, \pi$. 

The equations of motion for $\phi$ and $\phi^c$ are given by 
\bea
0 &=& 
\label{phieom}
\frac{W_b}{2M_5^3}
\left(
-2(\delta(y) - \delta(y-\pi))h^c 
+ \frac{\phi^\dag}{2M_5^3}W_b 
+ %\frac{7}{3}
7(\hat \epsilon(y))^2kR h^c - \hat \epsilon(y) \partial_y h^c
\right) \nonumber \\
&&
-R^2 e^{2R\sigma} \eta^{\mu\nu}\partial_\mu\partial_\nu \phi^\dagger
- e^{(\frac{5}{2}+c)R\sigma} \partial_y 
\left(
e^{-(1+2c)R\sigma} \partial_y (e^{-(\frac{3}{2}-c)R\sigma} \phi^\dag) 
\right), 
\eea
\bea
\label{phiceom}
0 &=& \frac{W_b}{2M^3}
\left( \frac{\hat \epsilon(y)h^{c\dag}}{2M^3} W_b  
+ 2\partial_y \phi \right) -2(2\delta(y) 
- 2\delta(y-\pi))\partial_y h^{c\dag}
\nonumber \\
&-&2(\partial_y \delta(y) - \partial_y 
\delta(y-\pi))h^{c\dag} 
+ \frac{\phi}{2M^3}\partial_y W_b \nonumber \\
&+&\hat \epsilon(y) \left[
-R^2e^{2R\sigma}
\eta^{\mu\nu}\partial_\mu\partial_\nu h^{c\dagger} 
- e^{(\frac{5}{2}-c)R\sigma} \partial_y 
\left(
e^{-(1-2c)R\sigma} 
\partial_y (e^{-(\frac{3}{2}+c)R\sigma} h^c{}^\dagger) 
\right)
\right] . \nonumber \\
\eea
One should note that the equations for $\phi$ and $\phi^c$ 
couple only through the boundary superpotential $W_b$. 
In the limit of vanishing boundary superpotential, we should 
obtain a series of effective fields with $\phi(x, y)$ component 
only, and another series of effective fields with $\phi^c(x, y)$ 
component only. 
For nonvanishing boundary superpotential, these two sets mix 
each other and we obtain 
$n$-th Kaluza-Klein effective field 
$\phi_n^I(x)$ with its mode functions $b_n^I(y)$ as $\phi(x, y)$ 
component and $b_n^{cI}(y)$ as $\phi^c(x, y)$ component 
\begin{eqnarray}
 \left(
  \begin{array}{c}
   \phi(x,y)\\ \phi^c(x,y)
  \end{array}\right)
  =\sum_{n}
\sum_{I=1,2}
 \phi_n^I(x) 
  \left(
   \begin{array}{c}
    b_n^I
(y)\\ \hat \epsilon(y)b_n^c{}^I
(y)
   \end{array}\right), 
\label{modeexpand}
\end{eqnarray}
where $I$ is the indices corresponding to the two independent 
effective fields eigenvalues. 

Assuming that the effective four-dimensional field 
$\phi_n(x)$ has mass $m_n$, 
we easily find solutions in the bulk in terms of the 
Bessel functions $J_\alpha, (J_\beta)$ and 
$Y_\alpha, (Y_\beta)$ 
\cite{GP}
\bea
b_n(y) &=& \frac{e^{2R\sigma}}{N_n}
\left[
J_\alpha(m_ne^{R\sigma}/k) 
+ b_\alpha(m_n)Y_\alpha(m_ne^{R\sigma}/k)
\right] , 
\quad \alpha =|c+\frac{1}{2}| , 
   \label{bn} \\
%\eea\bea
b_n^c(y) &=& \frac{e^{2R\sigma}}{N_n^c}
\left[
J_{\beta}(m_n e^{R\sigma}/k) 
+ b_{\beta}(m_n)Y_{\beta}(m_n e^{R\sigma}/k)
\right], 
\quad \beta=|c-\frac{1}{2}| , 
   \label{eq:bnc} 
\eea
where we have not yet specified which mode $I=1, 2$ until 
we determine mass eigenvalues later. 

Because of $\delta(y), \delta(y-\pi)$ in the boundary 
superpotential and also derivatives of sign function 
$\hat \epsilon(y)$, we obtain several types of singular 
contributions, $\delta^2(y)$ terms, $\partial_y\delta(y)$ 
terms and $\delta(y)$ terms and similarly for $y$ replaced 
by $y-\pi$. 
To find out these singular terms, we use the following 
identity valid as a result of a properly regularized 
calculation \footnote{
This follows from a wide range of regularization respecting 
the relation $2\delta(y)=d\epsilon(y)/dy$. 
}
\begin{equation}
\delta(y) (\hat \epsilon(y))^2 = \frac{1}{3} \delta(y), 
\quad 
\delta(y-\pi) (\hat \epsilon(y))^2 = \frac{1}{3} \delta(y-\pi). 
\label{eq:delta_epsilon2}
\end{equation}

All these singular terms need to be canceled resulting in 
boundary conditions. 
The first boundary condition comes from %cancelling 
$\delta^2$ terms in the equation of motion (\ref{phieom}) 
for $\phi$ 
\bea
\label{BC1}
-2 b_n^c(0) + w_0 b_n(0) &=& 0, \\
2 b_n^c(\pi) + w_\pi b_n(\pi) &=& 0. 
\label{BC2}
\eea
The second boundary condition comes from $\delta$ function 
in the equation of motion %(\ref{phieom}) 
for $\phi$ 
\bea
\label{BC3}
0 &=& \frac{7}{3}w_0 b_n^c(0) 
-\frac{1}{3}w_0 
\left[
2b_n^c(0) +\frac{m_n}{k} \frac{1}{N_n^c}
\left\{ J_{\beta}'(m_n/k) 
+ b_{\beta} Y_{\beta}'(m_n/k) \right\}
\right] \nonumber \\
&-&4b_n(0) 
+ 2 \left( \frac{3}{2}-c \right)b_n(0)
-\frac{2m_n}{k}
\left[ \frac{1}{N_n} 
\left\{ J_\alpha'(m_n/k) 
+ b_\alpha(m_n) Y_\alpha'(m_n/k) \right\} \right],
\\
0 &=& \frac{7}{3}w_\pi b_n^c(\pi) 
-\frac{1}{3}w_\pi 
\left[
2b_n^c(\pi) +\frac{m_n}{k} \frac{e^{3Rk\pi}}{N_n^c}
\left\{ J_{\beta}'(m_n e^{Rk\pi}/k) + 
b_{\beta} Y_{\beta}'(m_n e^{Rk\pi}/k) \right\} \right] 
\label{BC4}
%\nonumber 
\\
&+&4b_n(\pi) 
- 2 \left( \frac{3}{2}-c \right)b_n(\pi) 
+\frac{2m_n}{k}
\left[ \frac{e^{3Rk\pi}}{N_n} 
\left\{ J_\alpha'(m_n e^{Rk\pi}/k) 
+ b_\alpha(m_n) Y_\alpha'(m_n e^{Rk\pi}/k)\right\} \right], 
\nonumber 
%\\
\eea
with 
%$'$ denotes derivative with respect to $z$ such as 
$J'(z)=dJ(z)/dz$. 
We find that no additional boundary conditions 
\footnote{
Other potentially singular contributions do not contribute 
because of $\hat \epsilon \times \delta =0$, 
$\hat \epsilon \times \delta^2=0$. 
} 
arises 
from the equation of motion of $\phi^c$, 
since the boundary condition from 
$\partial_y \delta$ is identical to that from $\delta^2$. 

We wish to solve the boundary conditions 
(\ref{BC1})--(\ref{BC4}) in the limit 
\begin{equation}
m_n/k \ll 1, 
\label{eq:approx1}
\end{equation}
%\qquad
\begin{equation}
m_ne^{kR \pi}/k \gg 1. 
\label{eq:approx2}
\end{equation}
Because of (\ref{eq:approx1}), the coefficient $b_\alpha$ 
and $b_\beta$ is small and $Y_\alpha, Y_\beta$ terms 
can be neglected at $y=\pi$. 
As given in Appendix \ref{sc:bound_cod}, 
the condition (\ref{eq:approx2}) allows the remaining 
Bessel functions $J_\alpha$ and $J_\beta$ 
to be approximated by their asymptotic forms.

After changing variables, 
\bea
\frac{m_n}{k}e^{Rk\pi} \equiv x, 
\qquad 
\frac{2\alpha+1}{4}\pi \equiv a, 
\qquad 
\frac{2\beta+1}{4}\pi \equiv b,
\eea
we can rewrite boundary conditions at $y=\pi$ 
(\ref{apBC1}) and (\ref{apBC2}) as 
summarized in Appendix \ref{sc:bound_cod} into a matrix 
form 
\bea
\left[
\begin{array}{cc}
M_{11} & M_{12} \\
M_{21} & M_{22} %\\
\end{array}
\right]
\left[
\begin{array}{c}
1/N_n^c \\
1/N_n
\end{array}
\right] =0. 
 \label{matrix}
\eea
\begin{eqnarray}
  M_{11}&=&2 \cos(x-b), 
 \label{matrix11}
\\
  M_{12}&=&w_\pi \cos (x-a), 
 \label{matrix12}
\\
  M_{21}&=& {w_\pi \over 6} 
   \left((c^2-c+11) \cos(x-b) + 2x \sin(x-b) \right), 
 \label{matrix21}
\\
  M_{22}&=&\left(c(1-c)\cos (x-a) -2x \sin(x-a) \right).
 \label{matrix22}
\end{eqnarray}

Mass spectrum is determined by 
nontrivial solutions of fluctuations require the 
vanishing determinant of the matrix $M$ 
\bea
0 &=& 2\cos(x-b)[c(1-c)\cos(x-a) -2x \sin(x-a)] \nonumber \\
&&-w_\pi^2 \cos(x-a) 
\left( {c^2-c+11\over 6} \cos(x-b) + \frac{x}{3} \sin(x-b) \right), 
\label{det}
\eea
which determines the mass $m_n$. 
It is most useful to start from the limit of vanishing 
boundary superpotential, $w_\pi \rightarrow 0$, giving 
two possibilities. 
The first solution is purely $\phi(x,y)$ modes 
($\phi^c(x,y)=0$) 
\bea
0 = c(1-c)\cos(x-a) -2x \sin(x-a), 
\label{eq:phi_0_eq}
\eea
which gives solutions for large $x$ as 
\begin{equation}
x = n\pi +a + {c(1-c) \over 2n\pi} 
+ {\cal O}\left({1 \over n^2}\right), 
\label{eq:phi_0_sol}
\end{equation}
with integer $n$. 
The second solution is purely $\phi^c(x,y)$ modes 
($\phi(x,y)=0$) 
\bea
0 = \cos(x-b), 
\label{eq:phic_0_eq}
\eea
which gives solutions 
\begin{equation}
x = \left(n-{1\over 2}\right)\pi + b ,  
\label{eq:phic_0_sol}
\end{equation}
with integer $n$. 
When $w_\pi \not=0$, these two modes (\ref{eq:phi_0_sol}) 
and (\ref{eq:phic_0_sol}) mix each other to form 
two independent modes with slightly different mass spectra 
and with associated mode functions. 

We can rewrite the eigenvalue equation (\ref{det}) in a 
more convenient form 
\begin{eqnarray}
 \tan^2(x-a) - A(x) \tan (x-a) + B(x) =0, 
  \label{eq:tan_2nd_eq}
\end{eqnarray}
\begin{eqnarray}
A(x) \equiv {C\over 2x}
+\left(1+{w_\pi^2 \over 12} \right)\cot (a-b), 
\quad 
B(x) \equiv {C\over 2x}\cot(a-b) -{w_\pi^2\over 12},  
\label{eq:def_AB}
\end{eqnarray}
\begin{eqnarray}
C \equiv c(1-c)-{(c^2-c+11)w_\pi^2\over 12} . 
\label{eq:def_C}
\end{eqnarray}
Since we are interested in highly excited Kaluza-Klein 
states ($n \gg 1$), we can approximate $x \approx n\pi$ 
in the denominator of $C/(2x)$. 
Thus we find 
\begin{eqnarray}
 \tan (x-a) \approx 
{A(n\pi) \over 2}
\left( 1 \pm \sqrt{1- {4B(n\pi) \over A^2(n\pi)}}\right). 
 \label{eq:sol_gen}
\end{eqnarray}
We can determine the sign choice in (\ref{eq:sol_gen}) 
by examining the solution near $w_\pi \rightarrow 0$. 
Since the upper (lower) sign for $0 \le c \le 1$ 
($c< 0, 1< c$) reduces to the behavior in 
Eq.(\ref{eq:phi_0_sol}) at $w_\pi \rightarrow 0$, it should 
correspond to the mode reducing to the purely $\phi(x, y)$ 
component at $w_\pi \rightarrow 0$. 
We shall call this mode as the dominantly $\phi$ mode, and 
denote it by the mode functions ($I=1$) as 
$b_n^{1}(y), b_n^{c1}(y)$ and the effective field as 
$\phi_n^{I=1}(x) \equiv \phi_n(x)$ in equation 
(\ref{modeexpand}). 
The other choice of the sign corresponds to the mode 
 reducing to the purely $\phi^c(x, y)$ component 
at $w_\pi \rightarrow 0$ , 
which behaves as in Eq.(\ref{eq:phic_0_sol}). 
We shall call this mode as the dominantly $\phi^c$ mode, and 
denote the mode functions ($I=2$) as 
$b_n^{2}(y), b_n^{c 2}(y)$ and the effective field as 
$\phi_n^{I=2}(x) \equiv \phi_n^{c}(x)$ in equation 
(\ref{modeexpand}).

\vspace{5mm}

%\subsection
\noindent
\underline{\bf 
Dominantly $\phi$ mode: $\phi^{I=1} \equiv \phi$
}%\label{sc:phi_spectra}

\begin{figure}[h]
\begin{center}
\includegraphics[width=6.5cm]{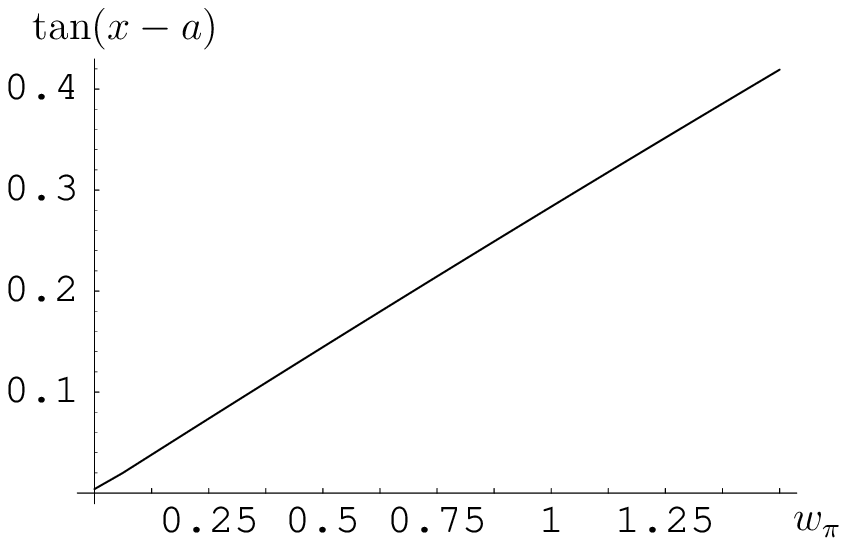}
\end{center}
\caption{The $w_\pi$ dependence of $\tan (x-a)$ with the 
upper sign (dominantly $\phi$ mode) in the case of $|c| \ge 1/2$. 
We use $c=0.5$ and $n=10$.\label{dxc1}} 
\end{figure}

We first consider the dominantly $\phi$ mode. 
Depending on the value of the bulk mass parameter $c$, 
there is a qualitative difference of mass spectrum 
as a function of $w_\pi$. 
Let us first consider the case of $|c|\ge 1/2$. 
Then the parameter becomes $a-b=\pm\pi/2$, 
which makes the energy levels for purely $\phi$ and $\phi^c$ 
modes to be split only small amount of the order of 
${\cal O}(1/n\pi)$ at high enough excitation ($n \gg 1$). 
Therefore we need to distinguish two cases depending on 
the relative magnitude of $|w_\pi|$ and $1/(n\pi)$. 
For $|w_\pi| \ll 1/(n\pi) \ll 1$, we obtain 
perturbation from nondegenerate eigenvalues to find 
a deviation proportional to $w_\pi^2$ 
\begin{eqnarray}
 \tan (x-a) \approx {c(1-c)\over 2n\pi}
\left[1+ \frac{1}{3}\left({n\pi w_\pi\over c(1-c)}\right)^2\right] .
\end{eqnarray}
For $w_\pi\sim {\cal O}(1)$, we obtain 
perturbation from (approximately) degenerate eigenvalues 
to find a deviation linear in $w_\pi$ 
\begin{eqnarray}
 \tan (x-a) \approx \pm {|w_\pi| \over 2\sqrt{3}}
  ~~\left\{
  \begin{array}{cll}
   + &\textrm{for}& 1/2\le c\le 1 \\
   - &\textrm{for}& c\le -1/2 ~\textrm{or}~ c>1
  \end{array}
   \right. , 
\end{eqnarray}
which gives the mass 
\begin{eqnarray}
 m_n 
\approx k e^{-Rk\pi}
  \left[ \left(n +{2\alpha+1\over 4} \right)\pi 
  \pm \frac{|w_\pi|}{2\sqrt{3}} \right], 
\quad w_\pi\sim {\cal O}(1) , 
%(n \gg 1) 
\label{constdx1}
\end{eqnarray}
where the plus (minus) sign should be taken for 
$1/2\le c \le 1 (c \le -1/2~\textrm{or}~c>1)$. \\
We show $\tan (x-a)$ with the upper sign 
(dominantly $\phi$ mode) as a function of $w_\pi$ in 
Fig.~\ref{dxc1}.

\begin{figure}[h]
\begin{center}
\includegraphics[width=6.5cm]{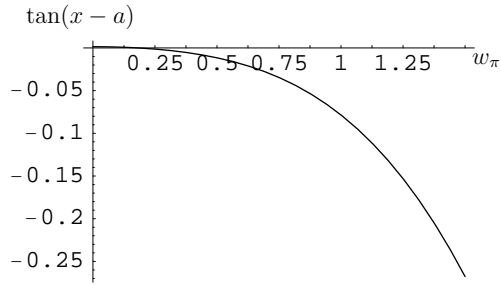}
\end{center}
\caption{
The $w_\pi$ dependence of $\tan (x-a)$ with the 
upper sign (dominantly $\phi$ mode) in the case of $|c| < 1/2$. 
We use $c=0.1$ and $n=10$.\label{dx1c2}} 
\end{figure}

For the case of $|c|<1/2$, the parameter becomes $a-b=c\pi$, 
which makes the energy levels for purely $\phi$ and $\phi^c$ 
modes to be nondegenerate even at high excitation ($n \gg 1$). 
Therefore we always obtain a mass shift proportional to 
$w_\pi^2$.  
For $w_\pi\ll 1/(n\pi)$, the solution is approximated by 
\begin{eqnarray}
 \tan (x-a) \approx 
{c(1-c)\over 2n\pi}-{w_\pi^2\tan c\pi\over 12}. 
\label{eq:tanx_smallc1}
\end{eqnarray}
For $w_\pi\sim {\cal O}(1)$, it is approximated by 
\begin{eqnarray}
 \tan (x-a) \approx 
{1\over 2\tan c\pi}\left(1+{w_\pi^2\over 12}\right)
 \left(1-\sqrt{1+{w_\pi^2\over 3}\tan^2 c\pi}\right) . 
\end{eqnarray}
This gives the mass whose $w_\pi$ dependence 
involves the bulk mass parameter $c$ 
\begin{eqnarray}
\label{phimass}
 m_n \approx k e^{-Rk\pi}
  \left[ \left(n +{2\alpha+1\over 4} \right)\pi 
+{w_\pi^2+12\over 24\tan c\pi}
   \left(1-\sqrt{1+{w_\pi^2\over 3}\tan^2 c\pi}\right) 
 \right].  
%\; (w_\pi\sim {\cal O}(1))
%\nonumber\\&& \label{phimass}
\end{eqnarray}
We show $\tan (x-a)$ as a function of $w_\pi$ 
in Fig.\ref{dx1c2}.

\vspace{5mm}

%\subsection
\noindent
\underline{\bf 
Dominantly $\phi^c$ mode: $\phi^{I=2} \equiv \phi^c$
}%\label{sc:phic_spectra}

\begin{figure}[h]
\begin{center}
\includegraphics[width=6.5cm]{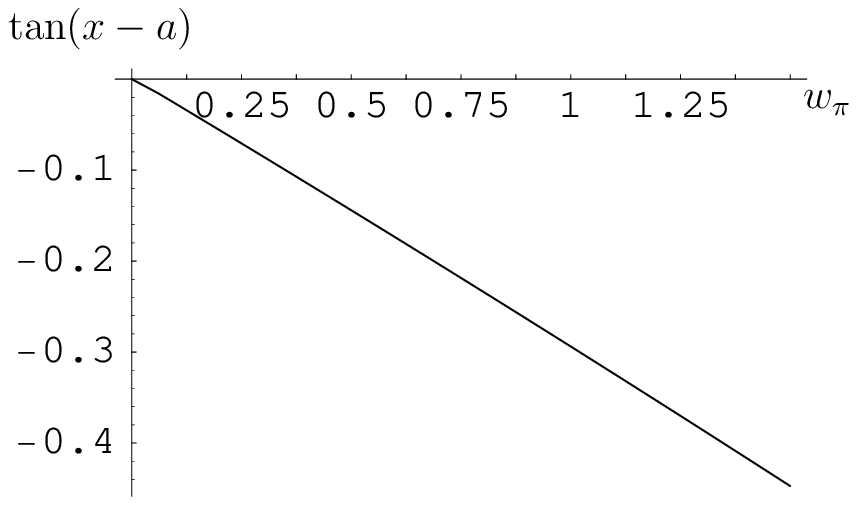}
\end{center}
\caption{The $w_\pi$ dependence of $\tan (x-a)$ with the 
lower sign (dominantly $\phi^c$ mode) 
in the case of $|c| \ge 1/2$. 
We use $c=0.5$ and $n=10$.
\label{dx2c1}} 
\end{figure}

If we choose the lower sign in Eq.(\ref{eq:sol_gen}), 
we obtain the dominantly $\phi^c$ mode. 
In the case of $|c| \ge 1/2$, we show 
%the solution is converted to 
$\tan (x-a)$ 
%(instead of $\tan(x-a)$) and is shown 
as a function of $w_\pi$ in Fig.\ref{dx2c1}. 
For $w_\pi\ll 1/(n\pi)\ll 1$ in the case of $|c| \ge 1/2$, 
we obtain
\begin{eqnarray}
   \tan (x-a)\approx -{n\pi\over 6c(1-c)}w_\pi^2. 
\label{eq:tanx_largec_smallw}
\end{eqnarray}
For $w_\pi \sim {\cal O}(1)$ in the case of $|c| \ge 1/2$, 
it becomes
\begin{eqnarray}
 \tan (x-a)\approx \mp{|w_\pi|\over 2\sqrt{3}} 
  ~~\left\{
  \begin{array}{cll}
   - &\textrm{for}& 1/2\le c\le 1 \\
   + &\textrm{for}& c\le -1/2 ~\textrm{or}~c>1
  \end{array}
   \right. , 
\label{eq:tanx_largec_largew}
\end{eqnarray}
which gives the mass 
\begin{eqnarray}
\label{phicmass}
 m_n 
&\approx& k e^{-Rk\pi}
  \left[ \left(n - \frac{1}{2} + {2\beta+1\over 4} \right)\pi 
  \mp \frac{|w_\pi|}{2\sqrt{3}} \right]~(n \gg 1). 
\label{constdx2}
\end{eqnarray}
where the plus (minus) sign should be chosen for 
$c \le -1/2~\textrm{or}~c>1(1/2\le c \le 1)$.

\begin{figure}[h]
\begin{center}
\includegraphics[width=6.5cm]{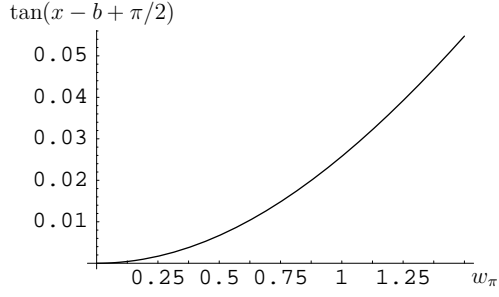}
\end{center}
\caption{The $w_\pi$ dependence of $\tan (x-b+\pi/2)$ for 
dominantly $\phi^c$ mode 
in the case of $|c| < 1/2$. 
We use 
$c=0.1$ and $n=10$.
\label{fig:tanx_phic_csmall}} 
\end{figure}

In the case of $|c|<1/2$, we 
show $\tan(x-b+\pi/2)$ instead of $\tan(x-a)$ in 
Fig.\ref{fig:tanx_phic_csmall}.
It is approximated as
\begin{eqnarray}
 \tan (x-b+\pi/2) \approx {w_\pi^2\tan c\pi \over 12} , 
\label{eq:tanx_smallc}
\end{eqnarray}
which gives the mass whose $w_\pi$ dependence involves 
the bulk mass parameter $c$ 
\begin{eqnarray}
 m_n \approx k e^{-Rk\pi}
  \left[ \left(n - \frac{1}{2} + {2\beta+1\over 4} \right)\pi 
  +{w_\pi^2\tan c\pi \over 12}\right] . 
\end{eqnarray}

As for the mass spectrum of hyperfermions, 
one can immediately check that the linearized equations 
of motion for hyperfermions are not affected by the 
constant boundary superpotentials $w_0, w_\pi$ 
localized at branes. 
Therefore the mass spectrum for hyperfermions is 
that for hyperscalars with $w_\pi=0$.

\setcounter{equation}{0}
%%%%%%%%%%%%%%%%%%%%%%%%
\section{Mode functions}\label{sc:mode_function}
%%%%%%%%%%%%%%%%%%%%%%%%
Let us determine the mode functions for each eigenvalue 
obtained above.
The corresponding eigenvector of each solution 
gives the ratio $N_n/N_n^c$ for the mode functions 
$(b_n,\hat \epsilon b_n^c)^T$ in 
Eqs.(\ref{modeexpand})--(\ref{eq:bnc}). 

\vspace{5mm}

%\subsection
\noindent
\underline{\bf 
Dominantly $\phi$ mode: $\phi^{I=1} \equiv \phi$
}%\label{sc:phi_spectra}

We first consider the dominantly $\phi$ modes corresponding 
to the choice of the upper sign in Eq.(\ref{eq:sol_gen}). 
Using (\ref{matrix})--(\ref{matrix22}), 
we find the eigenvector 
\bea
\frac{N_n}{N_n^c} = 
-\frac{M_{12}}{M_{11}} = -\frac{M_{22}}{M_{21}}, 
\eea
which is most easily evaluated by 
\begin{eqnarray}
 {M_{12}\over M_{11}}
%&
=%&
{w_\pi\over 2}
    {1\over \cos (a-b)-\tan (x-a) \sin (a-b)},
\label{eq:M12_M11}
%\\
\end{eqnarray}

\begin{figure}[h]
 \begin{center}
  \includegraphics[width=6.5cm]{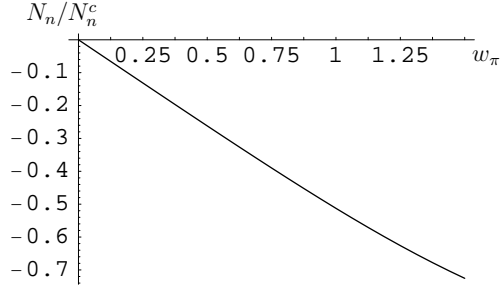}
 \end{center}
\caption{The ratio $N_n/N_n^c$ as a function $w_\pi$ 
in the case of $|c| < 1/2$. 
We use $c=0.1$ and $n=10$. \label{rat1c2}}
 \end{figure}

For $|c|\le 1/2$, 
we obtain the ratio $N_n/N_n^c$ which is shown as  a function 
of $w_\pi$ in Fig.\ref{rat1c2}, 
using the solution (\ref{eq:sol_gen}) with the upper sign 
for the dominantly $\phi$ modes. 
Using (\ref{eq:tanx_smallc1}) for small $w_\pi$, 
it can be approximated as 
\begin{eqnarray}
  {N_n\over N_n^c}\approx -{w_\pi\over 2\cos c\pi} .
\end{eqnarray}
This result represents ${\cal O}(w_\pi)$ mixing of 
$b_n^c$ as we expected.

\begin{figure}[h]
 \begin{center}
  \includegraphics[width=6.5cm]{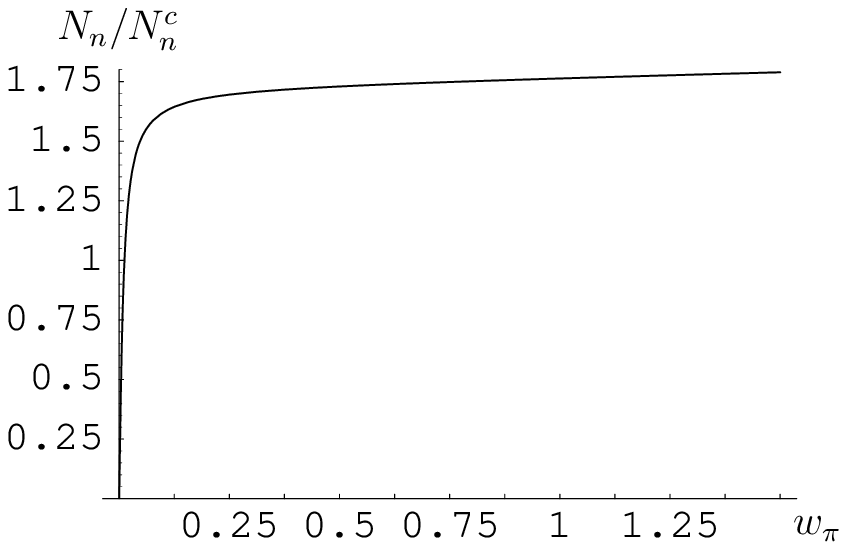}
 \end{center}
\caption{The ratio $N_n/N_n^c$ as a function $w_\pi$ 
for dominantly $\phi^c$ mode 
in the case of $|c| \ge 1/2$. 
We use 
$c=0.5$ and $n=10$. 
\label{rat1c1}}
 \end{figure}

For $|c|\ge 1/2$, we obtain the ratio by letting 
$a-b=\pm \pi/2$ in Eq.(\ref{eq:M12_M11}). 
Using the solution (\ref{eq:sol_gen}) with the 
upper sign for the dominantly $\phi$ modes, 
we show $N_n/N_n^c$ as a function of $w_\pi$ 
in Fig.\ref{rat1c1}. 
Using (\ref{eq:tanx_largec_smallw}) for 
$w\ll (1/n\pi)$, the ratio $N_n/N_n^c$ can be 
approximated by 
\begin{eqnarray}
 \left|{N_n\over N_n^c}\right|&\approx& 
 \left|{n\pi w_\pi \over c(1-c)}\right|.  
\end{eqnarray}
The result in this case represents ${\cal O}(w_\pi)$ mixing 
of $b_n^c$ characteristic of the perturbation between 
nondegenerate states. 
Using (\ref{eq:tanx_largec_largew}) for 
 $w_\pi\approx {\cal O}(1)$, we obtain a wave function 
with the 
${\cal O}(1)$ mixing between $\phi$ and $\phi^c$ 
components 
\begin{eqnarray}
 \left|{N_n\over N_n^c}\right| &\approx& \sqrt{3} , 
\end{eqnarray}
as anticipated by the $w_\pi$ perturbation between 
the nearly degenerate ($w_\pi \gg 1/n\pi$) states.

\vspace{5mm}

%\subsection
\noindent
\underline{\bf 
Dominantly $\phi^c$ mode: $\phi^{I=2} \equiv \phi^c$
}%\label{sc:phi_spectra}

Similarly we can obtain the dominantly $\phi^c$ modes 
corresponding to the choice of lower sign in 
Eq.(\ref{eq:sol_gen}). 
It is now more convenient to use the ratio 
\begin{eqnarray}
 {N_n^c\over N_n}
   =-{M_{21}\over M_{22}}, 
\end{eqnarray}
\begin{eqnarray}
&\!\!\!&\!\!\! {M_{21}\over M_{22}}={w_\pi\over 6} \times \\
&\!\!\!&\!\!\!\!\!\!\!\!\!
   {(c^2-c+11)\tan(x-b+{\pi\over 2})-2x\over 
  c(1-c)(\tan(x-b+{\pi\over 2})\cos(a-b)-\sin(a-b))
     +2x(\cos(a-b)+\tan(x-b+{\pi\over 2})\sin(a-b))} .
\nonumber 
\end{eqnarray}

\begin{figure}[h]
 \begin{center}
  \includegraphics[width=6.5cm]{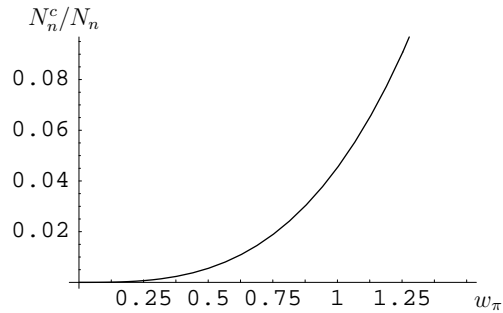}
 \end{center}
\caption{The ratio $N_n^c/N_n$ as a function $w_\pi$ 
for dominantly $\phi^c$ mode 
in the case of $|c| < 1/2$. 
We use 
$c=0.1$ and $n=10$. 
\label{rat2c2}}
 \end{figure}

For $|c|<1/2$ case, we show $N_n^c/N_n$ as a function of 
$w_\pi$ in Fig.\ref{rat2c2}, using the solution 
(\ref{eq:sol_gen}) with the lower sign for the dominantly 
$\phi^c$ modes. 
Using (\ref{eq:tanx_smallc}) for small $w_\pi$, the ratio 
is approximately given by 
\begin{eqnarray}
 {N_n^c\over N_n}\approx {w_\pi\over 6\cos c\pi} .
\end{eqnarray}
This result exhibits ${\cal O}(w_\pi)$ mixing of 
$\phi$ for the $w_\pi$ perturbation between 
nondegenerate states.

\begin{figure}[h]
 \begin{center}
  \includegraphics[width=6.5cm]{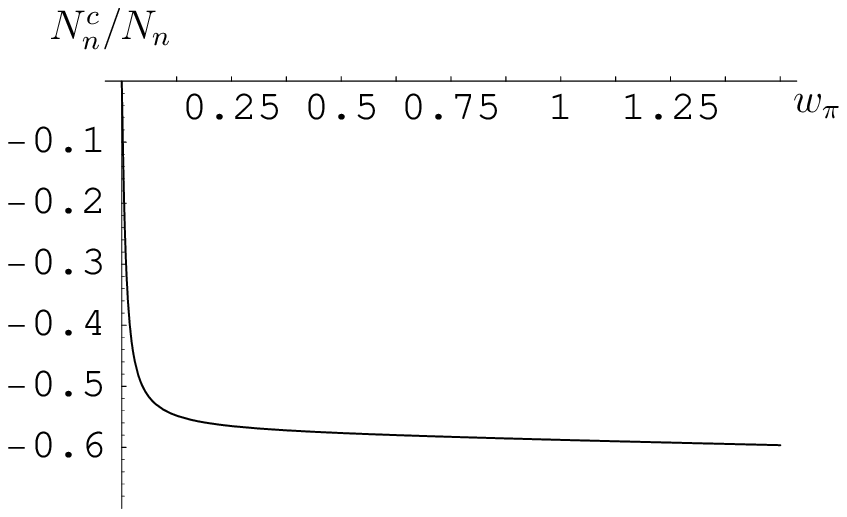}
 \end{center}
\caption{The ratio $N_n^c/N_n$ as a function $w_\pi$ 
for dominantly $\phi^c$ mode 
in the case of $|c| \ge 1/2$. 
We use 
$c=0.5$ and $n=10$. 
 \label{rat2c1}}
 \end{figure}

For $|c|\ge 1/2$, we show $N_n^c/N_n$ as a function of 
$w_\pi$ in Fig.\ref{rat2c1}, using the solution 
(\ref{eq:sol_gen}) with the lower sign for the dominantly 
$\phi^c$ modes. 
Using (\ref{eq:tanx_largec_smallw}) for 
$w_\pi \ll 1/(n\pi)$, the ratio 
is approximately given by 
\begin{eqnarray}
 \left|{N_n^c\over N_n}\right|&\approx& 
 \left|{n\pi w_\pi\over 3c(1-c)}\right|, 
\end{eqnarray}
which exhibits ${\cal O}(w_\pi)$ mixing of $\phi$ 
as a $w_\pi$ perturbation of nondegenerate states. 
Using (\ref{eq:tanx_largec_largew}) for 
$w_\pi\sim {\cal O}(1)$, 
we obtain 
${\cal O}(1)$ mixing between $\phi$ and $\phi^c$ 
components 
\begin{eqnarray}
 \left|{N_n^c\over N_n}\right|&\approx& {1\over \sqrt{3}}, 
\end{eqnarray}
as anticipated by the $w_\pi$ perturbation between 
the nearly degenerate ($w_\pi \gg 1/n\pi$) states.

\setcounter{equation}{0}
%%%%%%%%%%%%%%%%%%%%%%%%%%%%%%%%%%%%%%%%%%%%%
\section{Comment on the radius stabilization}\label{sc:radius_st}
%%%%%%%%%%%%%%%%%%%%%%%%%%%%%%%%%%%%%%%%%%%%%
In this section, we study the 
background solution for hypermultiplets and possible 
radius stabilization in the presence of 
constant boundary superpotential $W_b$ together with the 
compensator $\varphi$ and the radion $T$ supermultiplets.

As a simple and realistic approximation, 
we consider a perturbative treatment for small values 
of boundary superpotential $W_b$. 
Let us first find out a background solution for the case of 
vanishing boundary superpotential: $w_0=w_\pi=0$. 
By requiring no SUSY breaking with vanishing auxiliary fields 
$F=F^c=F_\varphi=F_T=0$ in Eqs.~(\ref{Feom})-(\ref{FTeom}), 
we obtain a unique supersymmetric solution with a complex 
constant $N_2$, %which can depend on $R$ 
%as the only arbitrary parameter 
\begin{eqnarray}
&& \phi=N_2\exp\left[\left({3\over 2}-c\right)R\sigma\right]
   \equiv \phi_s , \label{eq:susy_sol_w0}\\
&& \phi^c=0 \label{eq:susy_solc_w0}.
\end{eqnarray}
The potential vanishes for any values of the 
radius $R$ and the complex parameter $N_2$. 
Therefore the supersymmetric solution $\phi_s$ has a flat 
directions along $\phi$, as is typical for a SUSY solution. 
Moreover, radius $R$ is undetermined in this 
$w_0, w_\pi \to 0$ limit. 

For nonzero $w_0$ and $w_\pi$, we will find that 
supersymmetry is generically broken, and a nontrivial 
potential will be generated. 
Let us now assume $|w_0| \sim |w_\pi| \equiv w \ll 1$ and 
work out perturbative solutions of the equations of motion 
for $\chi$ and $\chi^c$ as deviations from the 
supersymmetric solutions in Eqs.~(\ref{eq:susy_sol_w0}) 
and (\ref{eq:susy_solc_w0})  
\begin{eqnarray}
   &&\phi=\phi_s+\chi, 
\label{eq:pert_sol}
\\
   &&\phi^c=\hat{\epsilon}\chi^c , 
\label{eq:pert_solc}
\end{eqnarray}
where we allow possible discontinuities of the $Z_2$ 
odd field $\phi^c$ across the fixed points $y=0, \pi$, 
similarly to Eq.~(\ref{eq:odd_field}). 
The equations of motion from the Lagrangian 
(\ref{aux}) shows that $\phi^c{}^\dagger$ arises in 
the first order in $w$ 
\begin{eqnarray}
 &&\left[\partial_y+\left({3\over 2}+c\right)R\sigma'\right]
 \left(e^{-4R\sigma}\left({\phi_s^\dagger\phi_s%-6M_5^3
%r_s
\over 6M_5^3}-1\right)
 \left[-\partial_y+\left({3\over 2}+c\right)R\sigma'\right]
\hat{\epsilon}\chi^c
 \right)
\nonumber\\
 && +e^{-4R\sigma}{\phi_s\phi_s^\dagger\over \phi_s^\dagger\phi_s-6M_5^3%r_s
}(3-2c)^2 (R\sigma')^2
    \hat{\epsilon}\chi^c
    -{e^{-4R\sigma}\over 2M_5^3 }\phi_s^\dagger (\partial_y W_b)=0 .
\label{eq:chic_EOM}
\end{eqnarray}
%where $r_s\equiv \phi_s^\dagger\phi_s-6M_5^3$. 
%We find that only the $\chi^c$ receives a contribution 
%to the first order of $W_b$. 
%

Provided the bulk fermion mass parameter $c \not = 3/2$, 
we can change a variable from $y$ to a dimensionless 
variable %$X\equiv r_s/(6M_5^3)$,
$X\equiv \phi_s^\dagger\phi_s/(6M_5^3)-1$, 
and obtain the field 
equation for the bulk 
($y \not =0, \pi$) 
\begin{eqnarray}
 &&\bigg[
   -(3-2c)^2X^2 (X+1)^2 {\partial^2\over \partial X^2}
\nonumber
\\
  && -(3-2c)X(X+1)(3-2c+2(1-2c)X){\partial\over \partial X}
\nonumber
\\
 && +(3-2c)^2+(3-2c)(9/2-c)X+(1/2-c)(3/2+c)X^2\bigg]
  \chi^c=0 , 
  \label{w1eqbulk}
\end{eqnarray}
whose solution is given for generic values of the bulk 
fermion mass parameter $c$ ($\not=3/2, 1/2$) as 
\begin{eqnarray} \label{solchic}
 \chi^c=
   X^{-1}(X+1)^{(5/2-c)/(3-2c)}
   \left[c_1 +c_2 (X+1)^{-(1-2c)/(3-2c)}(X+(3-2c)/(1-2c))\right],
\end{eqnarray}
where $c_1$ and $c_2$ are constants of integration. 
The solution for $c=1/2$ is given as 
\begin{eqnarray} \label{solchic2}
 \chi^c=
 X^{-1}(X+1) \left[c'_1 +c'_2 (X-\ln(X+1))\right], 
\end{eqnarray}
with another integration constants $c'_1, c'_2$. 
The field equation also yields the boundary conditions 
as the cancellation conditions of $\partial_y\delta(y)$ and 
$\partial_y\delta(y-\pi)$  
\begin{eqnarray} \label{bcchic}
 X(-\partial_y^2 \hat{\epsilon}) \chi^c
  -{1\over 2M_5^3}\phi_s^\dagger (\partial_y W_b)=0, 
\end{eqnarray}
respectively. 
Using $W_b$ 
in Eq.(\ref{eq:boundary_pot}) and 
the background solution $\phi_s$ in 
Eq.(\ref{eq:susy_sol_w0}), 
we obtain the boundary conditions more explicitly 
\begin{eqnarray}
 &&\chi^c\bigg|_{y=0}
  =-{\phi_s^\dagger w_0\over 2X}\bigg|_{y=0}
  ={N_2^\dagger w_0\over 2(1-\hat{N})} , 
\label{eq:bound_cond0}
\\
 &&\chi^c\bigg|_{y=\pi}
  ={\phi_s^\dagger w_\pi\over 2X}\bigg|_{y=\pi}
  ={N_2^\dagger w_\pi e^{(3/2-c)Rk\pi}
  \over 2(\hat{N} e^{(3-2c)Rk\pi}-1)} , 
\label{eq:bound_condpi}
\end{eqnarray}
where we defined a dimensionless parameter 
$\hat{N}\equiv |N_2|^2/(6M_5^3)$. 
These boundary conditions determine the integration constants 
$c_1, c_2$ (or $c'_1, c'_2$) in terms of the unique 
undetermined parameter $N_2$. 
We find for generic values of $c$ ($\not=1/2, 3/2$) 
\begin{eqnarray}
\!\!\! c_1 =
 -\left( \frac{N_2^\dag}{2\hat{N}^{\frac{5-2c}{2(3-2c)}}} \right) 
 \frac{((1-2c)\hat{N}e^{2Rk\pi}+2e^{-(1-2c)Rk\pi})w_0 
+ ((1-2c)\hat{N}+2)w_\pi e^{-Rk\pi}}
{(1-2c)\hat{N}(e^{2Rk\pi}-1)+2(e^{-(1-2c)Rk\pi}-1)}, 
\end{eqnarray}
%\\
\begin{eqnarray}
 c_2
%&\!\!\!&
=%&%&\!\!\!
\left({N_2^\dagger \over 2\hat{N}^{{3+2c \over 2(3-2c)}}}\right) 
  {(1-2c)  \left(w_0+w_\pi e^{-Rk\pi}\right) \over 
%\nonumber \\  &\!\!\!\times&\!\!\!
       %\left[
(1-2c)\hat{N}(e^{2Rk\pi}-1)
      +2(e^{-(1-2c)Rk\pi}-1)}
%\right]^{-1} 
. 
\label{eq:int_par}
\end{eqnarray}
whereas we find the solution for $c=1/2$ as 
\begin{eqnarray}
 c'_1&=& -\left( \frac{N_2^\dag}{2\hat{N}} \right)
     \frac{(\hat{N}e^{2Rk\pi}- \ln \hat{N} -1- 2Rk \pi) w_0 
     + (\hat{N} - \ln \hat{N} - 1 ) w_\pi e^{-Rk\pi}}
     {\hat{N}(e^{2Rk\pi}-1)-2Rk\pi}, \\
 c'_2&=&
   \left({N_2^\dagger \over 2\hat{N}}\right) 
  { w_0+w_\pi e^{-Rk\pi}\over %\left[
\hat{N}(e^{2Rk\pi}-1)-2Rk\pi}
%\right}]^{-1}
. 
\label{eq:int_par2}
\end{eqnarray}
In the case of $c=3/2$, we can solve Eq.(\ref{eq:chic_EOM}) 
in the bulk directly and find 
\begin{eqnarray} \label{solchic2}
 \chi^c=c''_1 e^{R\sigma} +c''_2 e^{3R\sigma}, 
\end{eqnarray}
whose integration constants $c''_1, c''_2$ are determined by 
the boundary condition as 
\begin{eqnarray} \label{solchic2}
 c''_1 &=& -\frac{N_2^\dag}{2(\hat{N}-1)}e^{2Rk\pi}
       \frac{(w_0 + w_\pi e^{-3Rk\pi})}{e^{2Rk\pi}-1}, \\
 c''_2 &=&  
   {N_2^\dagger \over 2(\hat{N}-1)} {
    \left(w_0+w_\pi e^{-Rk\pi}\right)
\over  e^{2Rk\pi}-1}. 
\end{eqnarray}

From the equation of motion for $\phi$, we find that 
the other perturbation $\chi$ of the background $\phi_s$ 
is of order ${\cal O}(w^2)$, and that 
it is given in terms of 
the background $\phi_s$ and the first order perturbation 
$\chi^c \sim {\cal O}(w)$ uniquely, except for the amount of the 
 admixture of the background solution $\phi_s$, which is always 
undetermined. 
Namely $\chi$ is determined in terms of $N_2$ 
%and $c_1, c_2$ (or $c'_1, c'_2$) 
without additional integration constants.

By inserting these solutions into the Lagrangian (\ref{aux}) and 
integrating over the extra dimension $y$, 
we obtain the potential as a function of the radius $R$ 
and the complex normalization parameter $N_2$ 
\begin{eqnarray}
&& V={k\over 2M_5^3}\int_0^{\pi}
   dy \bigg\{
  -2c_2^\dagger\hat{N}^{5/2-2c+2/(3-2c)}
     e^{((3-2c)(5/2-2c)+2)R\sigma}
\nonumber
\\
  &&\!\!\!\!\!\! +\left({3\over 2}+c+(3-2c)
  \left(-{5\over 2}+2c-\left[3(\hat{N}
 e^{(3-2c)R\sigma}-1)\right]^{-1}\right)\right)
   \chi^c{}^\dagger \bigg\}
    {\phi_s^\dagger W_b\over 2}e^{-4R\sigma} , 
\end{eqnarray}
where we should use the solution in Eq.(\ref{solchic}) 
subject to the boundary condition (\ref{bcchic}) for $\chi^c$. 
By performing integration and 
using the boundary conditions (\ref{eq:bound_cond0}) and 
(\ref{eq:bound_condpi}), 
we find the potential 
\begin{eqnarray}
 V&=& -N_2^\dagger k
  c_2^\dagger\hat{N}^{5/2-2c+2/(3-2c)}
     \left(w_0+w_\pi e^{((3-2c)^2-2)Rk\pi}\right)
\nonumber\\
  &&+{|N_2|^2 kw_0^2\over 4(1-\hat{N})}
       \left(-4c^2+12c-6+{3-2c\over 3(1-\hat{N})}\right)
\nonumber\\
 &&+{|N_2|^2 k w_\pi^2 e^{-(1+2c)Rk\pi}
     \over 4(\hat{N}e^{(3-2c)Rk\pi}-1)}
     \left(-4c^2+12c-6-{3-2c\over 3(\hat{N}e^{(3-2c)Rk\pi}-1)}\right) 
, 
 \label{potential123}
\end{eqnarray}
where we should use $c_2$ in Eq.(\ref{eq:int_par}) 
for generic values of $c$. 
It should be replaced by $c'_2$ in Eq.(\ref{eq:int_par2}) 
for $c=1/2$. 
We find that $c=1/2$ case can be obtained as a smooth 
limit from $c\not=1/2$. 
%, and $c_2\to 0$ for $c=3/2$. 
We see that the supersymmetry breaking induced by the 
boundary superpotential produces the 
potential $V$ as a nontrivial function of $N_2$ and $R$. 

\begin{figure}[h]
\begin{center}
 \includegraphics[width=7cm]{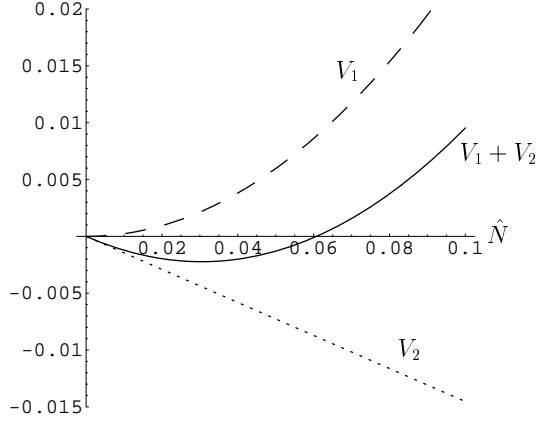}
 \caption{Potential for $c= \ccr-\Delta c$. \label{stabiliz}}
\end{center}
\end{figure}

Let us now study the stabilization of the radius $R$ and 
the modulus $N_2$. 
For simplicity, we consider the case where $w_\pi=0$ 
and the constant $N_2$ is real. 
Namely we assume that the source of the SUSY breaking 
is localized on only one of the brane at $y=0$ (Planck 
brane). 
Then the potential becomes
\bea
 V
&=& {3M_5^3 k w_0^2\over 2}\bigg\{
  \frac{-2(1-2c)}{(1-2c)(e^{2Rk\pi}-1)\hat{N} +2(e^{(2c-1)Rk\pi}-1)}
\hat{N}^{4-2c-\frac{1}{3-2c}} \nonumber \\
&&+\frac{\hat{N}}{1-\hat{N}}
\left( -4c^2+12c-6 +\frac{3-2c}{3(1-\hat{N})}
\right)
  \bigg\}. 
\label{potentialwp0}
\eea
%where we used $c_2$ given in Eq.(\ref{eq:int_par}).
We need to require the stationary condition for both 
modes $R$ and $N_2$ 
%As we have derived the equations as deviations from supersymmetric
%solutions with two zero modes corresponding to $R$ and $\hat{N}$,
%the model has two modes.
%The stationary conditions to examine are given by
\begin{eqnarray}
  {\partial V\over \partial R}=0 
  \textrm{~~and~~} 
 {\partial V\over \partial \hat{N}}=0 .
\end{eqnarray}
The former condition $\partial V/\partial R=0$ leads to 
\begin{eqnarray}
 \hat{N}=e^{-(3-2c)Rk\pi} ,
  \label{dvr0}
\end{eqnarray}
whereas the latter condition gives 
\begin{eqnarray}
 0&=&
  -{2\left(4-2c-{1\over 3-2c}\right)\hat{N}^{3-2c-{1\over 3-2c}}
  \over \hat{N}(e^{2Rk\pi}-1)-{2\over 2c-1}(e^{(2c-1)Rk\pi}-1)}
  +{2(e^{2Rk\pi}-1)\hat{N}^{4-2c-{1\over 3-2c}}
   \over \left[
 \hat{N}(e^{2Rk\pi}-1)-{2\over 2c-1}(e^{(2c-1)Rk\pi}-1)
  \right]^2}
\nonumber
\\
 &&+{1\over (1-\hat{N})^2}
   \left(-4c^2+12c-6+{(3-2c)(1+\hat{N})\over 3(1-\hat{N})}\right) .
   \label{dvn0}
\end{eqnarray}
The vacuum expectation values for $\hat{N}$ and $(Rk)$ are obtained 
as solutions of Eqs.(\ref{dvr0}) and (\ref{dvn0}),
depending only on 
the mass parameter $c$. 
We find that there is a unique nontrivial minimum with a finite 
value of the radius $R$ and the normalization $N_2$ for the 
flat direction $\phi$ provided 
$c < c_{\rm cr}$ with 
\begin{eqnarray}
  c_{\textrm{\scriptsize cr}}\equiv
    {17-\sqrt{109}\over 12} .
\end{eqnarray}
At the critical value of the mass parameter $c_{\rm cr}$, 
the minimum occurs at infinite radius and vanishing 
normalization $N_2$ 
%One solution is 
\begin{eqnarray}
   \hat{N}(\ccr)=0 ,~~ R(\ccr)=\infty . 
%,~~    c_{\textrm{\scriptsize cr}}\equiv {17-\sqrt{109}\over 12} .
\end{eqnarray}
%This solution implies that the radius is stabilized at
%$R=\infty$ for $c=\ccr$.
To examine the stabilization for $c < c_{\rm cr}$ more closely, 
%at a finite radius,
we parametrize $c=\ccr-\Delta c$ with a small $\Delta c$. 
After using the relation Eq.(\ref{dvr0}), 
$\hat{N}=e^{-(3-2c)Rk\pi}$, 
we find that the potential (\ref{potentialwp0}) for 
$c=\ccr-\Delta c$ at the leading order of 
$\Delta c$ and $\hat{N}$ consists of two pieces 
\begin{eqnarray}
  V  &\approx& {3M_5^3 kw_0^2\over 2}(V_1 +V_2) ,
\\
  V_1&\equiv& {2(2\ccr-1)\over 3-2\ccr}
    \hat{N}^{4\ccr^2-12\ccr+10\over 3-2\ccr} , 
\\
  V_2&\equiv& -\hat{N}\left(-8\ccr+{34\over 3}\right)\Delta c 
. 
\end{eqnarray}
The first piece $V_1$ is positive and increases with 
a power larger than unity as a function of $\hat N$, 
whereas the second piece is negative and linear in 
$\hat N$, whose coefficient is proportional to $\Delta c$. 
The potential $V$ and its pieces $V_1, V_2$ are depicted 
as a function of $\hat N$ 
in Fig.\ref{stabiliz}. 
%The parts of the potential $V_1$ and $V_2$ are positive and
%negative, respectively.
It is now obvious that a unique minimum occurs at 
finite values of $\hat N$ provided $\Delta c >0$ 
($c < c_{\rm cr}$) and that the minimum point approaches 
$\hat N \to 0$ as $\Delta c \to 0$ 
($c \to c_{\rm cr}$).  
Actually the Fig.\ref{stabiliz} demonstrates only 
the stability along the direction of $\hat N$, 
after the other variable $R$ is eliminated by the 
stationary condition (\ref{dvr0}). 
We have checked that this minimum point gives 
a true minimum of the potential $V(R, \hat N)$ 
as a function of two variables, establishing the stability 
in both directions. 
%These balance yields a minimum of the potential.
%The slop of $V_2$ depends on $\Delta c$.
%Hence the stationary point is shifted depending on the size 
%of $\Delta c$.
%As shown above, for $\Delta c=0$, the stationary point is at 
%$\hat{N}=0$.
For $\Delta c\neq 0$,
the stationary condition $\partial V/\partial \hat{N}=0$
becomes
\begin{eqnarray}
  0\approx
   {2(2\ccr-1)\over 3-2\ccr}
  \left(3-2\ccr
   +{1\over 3-2\ccr}\right)
   \hat{N}^{2-2\ccr+{1\over 3-2\ccr}}
  -\left({34\over 3}-8\ccr\right)\Delta c .
\end{eqnarray}
From this equation, we find that the stationary point at 
the leading order of $\Delta c$ as 
\begin{eqnarray}
 && R\approx
    {-1\over \left[2(1-\ccr)(3-2\ccr)+1\right]k\pi}
      \ln \left[
     {(3-2\ccr)\left({17\over 3}-4\ccr\right)
     \over 2(2\ccr-1)\left(2-\ccr-{1-\ccr\over 3-2\ccr}\right)}
   \Delta c\right] ,
\\
 && 
  \hat{N}\approx
   \left[
   {(3-2\ccr)\left({17\over 3}-4\ccr\right)
   \over 2(2\ccr-1)\left(2-\ccr-{1-\ccr\over 3-2\ccr}\right)}
   \Delta c\right]^{3-2\ccr\over (3-2\ccr)(2-2\ccr)+1} .
\end{eqnarray}
Numerically $R$ is evaluated as
\begin{eqnarray}
  R\approx {1\over 10k}\left(\ln {1\over \Delta c}-3.4\right) .
\end{eqnarray}
which implies that $Rk>1$ is satisfied for $\Delta c<10^{-6}$. 
Therefore our model with supersymmetry breaking by 
constant boundary superpotentials realizes 
the radius stabilization at $Rk>1$ without requiring
any additional mechanism. 

Another attempt for a possible approximation is given 
in Appendix \ref{sc:background}.

%%%%%%%%%%%%%%%%%%%%
\section{Conclusion}\label{sc:conclusion}
%%%%%%%%%%%%%%%%%%%%

In this paper, we have studied SUSY breaking by constant 
superpotentials at the boundaries of the %warped 
Randall-Sundrum geometry. 
We have calculated the Kaluza-Klein mass 
spectrum of hypermultiplet in the SUSY %warped space 
Randall-Sundrum 
model including constant superpotentials 
localized at branes. 
In particular, we have taken account of 
the bulk mass parameter $c$, radion and compensator.

Here we note that the presence of the compensator does 
contribute to the above calculated spectrum of hyperscalars. 
In the auxiliary field Lagrangian (\ref{auxap}),
$F$ component of the compensator gives the following type 
of terms 
\begin{eqnarray}
      W_b \phi \partial_y\phi^c ,~~
      W_b \phi^c \partial_y\phi ,~~
      W_b \sigma' \phi \phi^c ,~~
      W_b^2 (\phi^\dagger\phi +\phi^c{}^\dagger\phi^c),
   \label{cmpsaterms}
\end{eqnarray}
and their hermitian conjugate. 
From (\ref{cmpsaterms}), the equation of motion for 
$\phi$ includes 
$%\begin{eqnarray}
 W_b \partial_y \phi^c , ~~
 W_b^2 \phi^\dagger , 
$ %   \label{cmpsateom}\end{eqnarray}
which are relevant to the boundary condition coming from 
the $\delta^2$ terms as in (\ref{BC2}) and 
$%\begin{eqnarray}
 W_b \partial_y \phi^c , ~~
 W_b \sigma' \phi^c ,
$ %  \label{cmpsateom2}\end{eqnarray}
which are relevant to the boundary condition coming from 
the $\delta$ terms as in (\ref{BC4}).
Therefore the boundary conditions (\ref{matrix}) to 
determine mass eigenvalues is different 
if we do not introduce the compensator.

In addition, we find that the SUSY breaking mass splitting 
and mode functions exhibit qualitatively different $w_\pi$ 
dependence for different values of the bulk mass parameter $c$.

As mentioned in Introduction, 
it is not yet clear what is the SS SUSY breaking 
in warped space. 
However, it has been shown that the SS SUSY breaking and 
SUSY breaking by the 
brane localized constant superpotentials are equivalent 
in flat space \cite{MP, BFZ, GR}. 
Therefore our results should reduce to the SS 
SUSY breaking if we take the flat limit. 
Since we have used the approximation $m_n/k \ll 1$ in 
order to focus on the effect of the warp factor, 
we cannot take the flat limit immediately from our results. 
Nevertheless it is interesting to observe the following 
similarity with the SS SUSY breaking in our mass spectrum. 
Let us first note that 
the bulk mass parameter $c$ should have large magnitude 
in order to take a proper flat limit $k\to 0$ as seen 
from Eq.(\ref{lagrangian}). 
Our result for large $|c|$ shows a linear dependence 
on $w_\pi$, which is quite 
similar to that of the SS SUSY breaking as summarized 
in Appendix \ref{sc:SSbreaking}.
This similarity seems to suggest 
the equivalence of the SS SUSY breaking with the 
SUSY breaking by constant superpotentials might be 
extendable in some way from flat space to the warped 
space. 
This is in accord with a recent proposal that 
the SS mechanism works in the supergravity 
gauged by a $Z_2$ even coupling \cite{AS}. 
On the other hand, we have found that the mass splitting 
depends on the bulk mass parameter 
for $|c| <1/2$, which is a new pattern of SUSY breaking. 
It would be interesting that the different behavior for 
different values of the bulk mass parameter $c$ might 
be understood by the analysis in the gauged supergravity. 
We hope that this work would give some insights to clarify 
whether the equivalence holds between SS breaking and SUSY 
breaking by constant superpotentials even in warped space.

We have also discussed the possibility of radius 
stabilization. 
In the limit of 
vanishing constant boundary superpotentials, 
we have obtained a SUSY background solutions as a solution of 
the classical equations of motion. 
The solution has two flat directions, the radius $R$ and 
the complex moduli parameter $N_2$ for the amplitude 
along one of the hypermultiplet scalar $\phi$. 
Then, we have introduced constant boundary superpotentials 
$w_{0,\pi} \ll 1$ as small perturbations. 
For the simplest situation of $w_{\pi}=0$ (boundary 
superpotential only for the Planck brane), 
we obtain the deviations from the SUSY solutions 
in the leading order of $w_{0}$. 
Using these solutions, 
we have explicitly calculated the radion potential, 
and have shown that the radius $R$ and the moduli 
$N_2$ are indeed stabilized at finite values. 
%In our approximation, the classical solution has an 
%undetermined overall factor which can depend on the 
%radius. 
%Therefore the radion potential cannot be completely 
%determined, preventing us to make a definite statement. 
%To improve this situation, we have to include the next 
%leading terms in the potential which are quite complicated. 
%Moreover, we need some additional mechanism to prevent 
%$\phi, \phi^c$ fields to become of order the Planck scale 
%${\cal O}(M^{3/2})$. 
%Otherwise we have to abandon our approximation 
%(\ref{eq:apprx1}), (\ref{eq:apprx2}). 

When the classical solution is trivial $\phi=\phi^c=0$, 
the radion potential is generated 
at 1-loop after SUSY is broken and 
the radius might be stabilized by the Casimir energy. 
If SUSY is broken at high energy, the radion is stabilized 
for a special bulk mass parameter $c=1/2$ \cite{GaP}. 
This mechanism might be helpful in the present case.

The radius stabilization has been studied also in the 
AdS$_4$ background where SS SUSY breaking can be formulated. 
In models with nonzero superpotential \cite{Katz:2006mv}, 
it has been found that hypermultiplets give positive 
contributions to the radion potential, contrary to the 
negative contributions from the gravity multiplet. 
This provides various patterns of radion potential. 
However explicit calculation with arbitrary values of $c$ 
remains to be studied. 
It would be interesting to apply our analysis for arbitrary 
values of $c$ to the radius stabilization of the AdS$_4$ 
background.

\vspace*{10mm}
%%%%%%%%%%%%%%%%%%%%
\begin{center}
{\bf Acknowledgements}
\end{center}  
The authors thank Y.~Sakamura for a useful discussion and 
Yuri Shirman for bringing interesting works to our attention. 
We also thank a referee for a valuable comment on the radius stabilization. 
This work is supported in part by Grant-in-Aid for Scientific 
Research from the Ministry of Education, Culture, Sports, 
Science and Technology, Japan No.17540237 (N.S.) 
and 16028203 for the priority area ``origin of mass'' 
(N.S.~and N.U.). 
N.M. is supported by INFN.

%\newpage
\begin{appendix}
 
\setcounter{equation}{0}
%%%%%%%%%%%%%%%%%%%%%%%%%%%%%%%%%%%%%%%%%%%%%%%%%%%%%%%%%
\section{Solving boundary conditions for mass spectrum}
%%%%%%%%%%%%%%%%%%%%%%%%%%%%%%%%%%%%%%%%%%%%%%%%%%%%%%%%%
\label{sc:bound_cod}

The boundary conditions from $\delta^2$ terms 
(\ref{BC1}) and (\ref{BC2}) can be rewritten 
in terms of the 
Bessel functions 
\bea
\label{BC11}
%&&
\frac{1}{N_n}\left[ J_\alpha(m_n/k) + b_\alpha(m_n) 
Y_\alpha(m_n/k) \right] 
= \frac{2}{w_0N_n^c} \left[ J_{\beta}(m_n/k) + b_{\beta}(m_n) 
Y_{\beta}(m_n/k) \right], 
%\\
\eea
\bea
&&
\frac{1}{N_n}\left[ J_\alpha(m_ne^{Rk\pi}/k) + b_\alpha(m_n) 
Y_\alpha(m_n e^{Rk\pi}/k) \right] \nonumber \\
&&\hspace*{30mm}= -\frac{2}{w_\pi N_n^c} 
\left[ J_{\beta}(m_ne^{Rk\pi}/k) 
+ b_{\beta}(m_n) Y_{\beta}(m_n e^{Rk\pi}/k) \right]. 
\label{BC21}
\eea
Similarly, the boundary conditions from $\delta$ terms 
(\ref{BC3}) and (\ref{BC4}) can be rewritten 
\bea
\label{BC31}
&&\left( \frac{5}{6}w_0^2-1-2c \right)
\left( J_\alpha(m_n/k) + b_\alpha(m_n) Y_\alpha(m_n/k) \right)
\nonumber \\
&&=
\frac{2m_n}{k} \left( 
J'_\alpha(m_n/k) + b_\alpha(m_n) Y'_\alpha(m_n/k)
\right) 
+\frac{1}{3}w_0 \frac{m_n}{k} \frac{N_n}{N_n^c}
\left[
J_{\beta}'(m_n/k) + b_{\beta}(m_n) Y_{\beta}'(m_n/k)
\right], \nonumber \\
\\
&&\left( \frac{5}{6}w_\pi^2-1-2c \right)
\left( J_\alpha(m_ne^{Rk\pi}/k) 
+ b_\alpha(m_n) Y_\alpha(m_ne^{Rk\pi}/k) 
\right) \nonumber \\
&&=
\frac{2m_n}{k} e^{Rk\pi} \left( 
J'_\alpha(m_ne^{Rk\pi}/k) 
+ b_\alpha(m_n) Y'_\alpha(m_ne^{Rk\pi}/k)
\right) \nonumber \\
&&-\frac{1}{3}w_\pi \frac{m_n}{k} \frac{N_n}{N_n^c}
\left[ e^{Rk\pi}
(J_{\beta}'(m_ne^{Rk\pi}/k) 
+ b_{\beta}(m_n) Y_{\beta}'(m_ne^{Rk\pi}/k))
\right]. 
\label{BC41}
\eea

We can determine $b_\alpha, b_\beta$ 
from the boundary conditions (\ref{BC11}) and (\ref{BC31}) 
at $y=0$ 
\bea
b_\alpha(m_n) 
&=& 
\frac{1}{(\beta+5)w_0^2 - 6( 1+ 2c - 2\alpha)} \nonumber \\
&&\times\left[
  \left(-(\beta+5)w_0^2+6(1+2c+2\alpha) \right)
  {J_\alpha(m_n/k)\over Y_\alpha(m_n/k)}
  +4 \beta w_0\frac{N_n}{N_n^c} 
{J_\beta(m_n/k)\over Y_\alpha(m_n/k)} 
  \right] 
\nonumber\\
&\sim&
 \frac{1}{(\beta+5)w_0^2 - 6( 1+ 2c - 2\alpha)} 
\nonumber \\
&& \times \left[
  \left((\beta+5)w_0^2-6(1+2c+2\alpha) \right)
  \left( \frac{m_n}{2k} \right)^{2\alpha}
  \frac{\pi}{\Gamma(\alpha+1)\Gamma(\alpha)} \right. 
\nonumber \\
&& \left. -4 \beta w_0\frac{N_n}{N_n^c}  
  \left( \frac{m_n}{2k} \right)^{\alpha+\beta}
  \frac{\pi}{\Gamma(\beta+1)\Gamma(\alpha)}
  \right],
 \label{balpha}\\
b_\beta(m_n) 
&=&   \frac{1}{(\beta+5)w_0^2 - 6( 1+ 2c - 2\alpha)}
\nonumber \\
&&\times\left[ 12\alpha w_0 
{J_\alpha(m_n/k)\over Y_\beta(m_n/k)} 
{N_n^c\over N_n}
  +\left((\beta-5)w_0^2 +6(1+2c-2\alpha)
    \right){J_\beta(m_n/k)\over Y_\beta(m_n/k)} \right] 
\nonumber\\
&\sim&
  \frac{1}{(\beta+5)w_0^2 - 6( 1+ 2c - 2\alpha)}
\nonumber \\
&&\times\left[ -12\alpha w_0  
\left( \frac{m_n}{2k} \right)^{\alpha+\beta}
  \frac{\pi}{\Gamma(\alpha+1)\Gamma(\beta)}
{N_n^c\over N_n} \right. \nonumber \\
&& \left.  -\left((\beta-5)w_0^2 +6(1+2c-2\alpha)
    \right) 
    \left( \frac{m_n}{2k} \right)^{2\beta}
  \frac{\pi}{\Gamma(\beta+1)\Gamma(\beta)}
    \right],
\label{bbeta}
\eea
where we used the approximation for $|z| \ll 1$
\bea
\frac{J_\beta(z)}{Y_\alpha(z)} 
\sim \left( \frac{z}{2} \right)^{\beta}
\left( -\frac{1}{\Gamma(\beta+1)} \right) 
\left( \frac{z}{2} \right)^{\alpha}
\sin (\pi \alpha) \Gamma(1-\alpha)
= -\left( \frac{z}{2} \right)^{\alpha+\beta}
\frac{\pi}{\Gamma(\beta+1)\Gamma(\alpha)}. 
\eea
In the approximation (\ref{eq:approx1}), 
we can neglect $b_\alpha, b_\beta$ since 
$(m_n/k)^{2\alpha}, (m_n/k)^{\alpha+\beta}, (m_n/k)^{2\beta}$ 
are very small. 
Using the other approximation condition (\ref{eq:approx2}), 
we can approximate the Bessel functions for $|z| >> 1$ as 
\bea
&&J_\alpha(z) \sim \sqrt{\frac{2}{\pi z}} 
\left[\cos \left( z - \frac{2\alpha+1}{4}\pi \right) 
 -{4\alpha^2-1\over 8z}\sin\left( z - \frac{2\alpha+1}{4}\pi \right) 
 \right], \\
&&Y_\alpha(z) \sim \sqrt{\frac{2}{\pi z}} 
\left[\sin \left( z - \frac{2\alpha+1}{4}\pi \right)  
-{4\alpha^2-1\over 8z}\sin\left( z - \frac{2\alpha+1}{4}\pi \right) 
 \right],
\eea
Therefore, it is a good approximation to impose the 
boundary conditions 
at $y=\pi$ with only $J_{\alpha(\beta)}$ terms 
disregarding $Y_{\alpha(\beta)}$ terms. 
Then, the approximated boundary conditions at 
$y=\pi$ (\ref{BC3}) and (\ref{BC4}) can be read 
\bea
\label{apBC1}
0 &\sim&
\frac{2}{N_n^c}
\cos(\frac{m_n}{k}e^{Rk\pi} - \frac{2\beta+1}{4}\pi)
+\frac{w_\pi}{N_n} 
\cos(\frac{m_n}{k}e^{Rk\pi} - \frac{2\alpha+1}{4}\pi), \\
0 &\sim& \frac{1}{N_n}
\left[c(1-c) \cos(\frac{m_n}{k}e^{Rk\pi} - \frac{2\alpha+1}{4}\pi) 
- \frac{2m_n}{k} e^{Rk\pi}
\sin(\frac{m_n}{k}e^{Rk\pi} - \frac{2\alpha+1}{4}\pi) 
\right] \nonumber \\
&&+ \frac{w_\pi}{N_n^c}
\left[ {c^2-c+11\over 6} 
\cos(\frac{m_n}{k}e^{Rk\pi} - \frac{2\beta+1}{4}\pi)
+ \frac{m_n e^{Rk\pi}}{3k} 
\sin(\frac{m_n}{k}e^{Rk\pi} - \frac{2\beta+1}{4}\pi) \right]
\nonumber
\\
\label{apBC2}
\eea
which gives the matrix eigenvalue equation (\ref{matrix}).

\setcounter{equation}{0}
%%%%%%%%%%%%%%%%%%%%%%%%%%%%%%%%%%%%%%%%%%%%%%%%%%%%%%%%%
\section{Relation to SS SUSY breaking}\label{sc:SSbreaking}
%%%%%%%%%%%%%%%%%%%%%%%%%%%%%%%%%%%%%%%%%%%%%%%%%%%%%%%%%
In this appendix, 
we briefly review that 
the SS SUSY breaking and SUSY breaking 
by the brane localized constant superpotentials are 
equivalent in flat space \cite{MP, BFZ, GR} and 
compare our spectrum in the SUSY Randall-Sundrum model 
with that in flat space. 
Let us consider the following hypermultiplet 
Lagrangian in flat space 
\bea
{\cal L} = \int d^4\theta (|\Phi|^2 + |\Phi^c|^2) 
+\left[ \int d^2\theta \Phi^c \partial_y \Phi +{\rm h.c.} \right].
\eea
Its auxiliary field Lagrangian can be read as
\bea
{\cal L}_{{\rm aux}} = (|F|^2+|F^c|^2) 
+ \left[
F^c \partial_y \phi
-F \partial_y \phi^c +{\rm h.c.}
\right],
  \label{flatal}
\eea
then, it is easy to solve 
\bea
F^\dag = \partial_y \phi^c, \quad
F^{c\dag} = - \partial_y \phi.
\eea
Here, we twist the hyperscalar in terms of 
R-symmetry rotation $SU(2)_R$, 
%$SU(2)_R \times U(1)_H$ rotations where $SU(2)_R$ is an R-symmetry 
%and $U(1)_H$ is just a phase rotation like a $U(1)_A$ symmetry, 
\bea
\left(
\begin{array}{c}
\phi \\
\phi^{c\dag}
\end{array}
\right)(y+2\pi R) = 
e^{2\pi i q\sigma_2} %(q \sigma_2 + \beta \sigma_3)}
\left(
\begin{array}{c}
\phi \\
\phi^{c\dag}
\end{array}
\right)(y), 
\label{twist}
\eea
where $q$ %, \beta$ are $SU(2)_R, U(1)_H$ charges. 
is $SU(2)_R$ charge.
This can be written as
\bea
\left(
\begin{array}{c}
\phi \\
\phi^{c\dag}
\end{array}
\right)(y) = 
e^{i\sigma_2 f(y)} %(\sigma_2 f(y) + \beta \sigma_3 y)}
\left(
\begin{array}{c}
\tilde{\phi} \\
\tilde{\phi}^{c\dag}
\end{array}
\right)(y), 
\label{redefine}
\eea
where $\tilde{\phi}^c$ are periodic functions for $S^1$ and 
$f(y)$
was found in \cite{BFZ}, 
\bea
f(y) =\frac{w_0-w_\pi}{4}\hat \epsilon(y) + \frac{w_0+w_\pi}{4}\eta(y)
\eea
where
\bea
\eta(y) = 2l+1, \quad l \pi R  < y < (l+1)\pi R , (l:{\rm integer})
\eea
is the ``staircase'' function that jumps by two units 
for every $\pi R$ along $y$. 
Note $f(y+2\pi R) = f(y) + w_0 + w_\pi$, which implies 
$w_0 + w_\pi =2\pi q$ to yield a correct twist. 
Then the equation of motion for $\tilde{\phi}$ is
\begin{eqnarray}
 (-R^2\eta^{\mu\nu}\partial_\mu\partial_\nu-\partial_y^2)\tilde{\phi}
 -\partial_y(f'\tilde{\phi}^c{}^\dagger)
 -f'(\partial_y\tilde{\phi}^c{}^\dagger-f'\tilde{\phi})=0 .
\end{eqnarray}
The equation of motion for $\tilde{\phi}^c{}$ is given by
\begin{eqnarray}
 (-R^2\eta^{\mu\nu}\partial_\mu\partial_\nu-\partial_y^2)\tilde{\phi}^c
   +\partial_y(f'\tilde{\phi}^\dagger)
  +f'(\partial_y\tilde{\phi}^\dagger+f'\tilde{\phi}^c)=0 .
\end{eqnarray} 
These equations can be solved by 
\begin{eqnarray}
  &&\tilde{\phi}=\cos(ny+\tilde{f}) 
\\
  &&\tilde{\phi}^c=\sin(ny+\tilde{f})
\end{eqnarray}
where $\tilde{f}(y)$ is the periodic function given by 
\begin{eqnarray}
 && \tilde{f}=\left\{
   \begin{array}{lll}
    {w_0\over 2}+{w_0+w_\pi\over 2\pi}(2l\pi-y)  
                  &\textrm{for}& 2l\pi <y<(2l+1)\pi\\
    {w_\pi\over 2}+{w_0+w_\pi\over 2\pi}((2l+1)\pi-y) &\textrm{for}&
     (2l+1)\pi<y<2(l+1)\pi \\
    0 &\textrm{for}& y=l\pi \\
   \end{array}\right.
\\
 && \tilde{f}'=w_0\delta(y)+w_\pi\delta(y-\pi)-{w_0+w_\pi\over 2\pi} .
\end{eqnarray}
The mass spectrum is obtained as\footnote{
Here, a regularization different from ours in 
(\ref{eq:delta_epsilon2}) 
is used for $\delta(y)\epsilon^2(y)$ by taking the limit for 
$\epsilon(y)$ first without respecting the relation 
$2\delta(y)=d\epsilon(y)/dy$. 
This treatment results in $\delta(y)\epsilon^2(y)=\delta(y)$, 
instead of our $\delta(y)\epsilon^2(y)=\delta(y)/3$ as 
in (\ref{eq:delta_epsilon2}). 
}
\begin{eqnarray}
  m_n={1\over R}\left[n-{w_0+w_\pi\over 2\pi}\right] .
\end{eqnarray}
This spectrum in flat case is similar to our results 
(\ref{constdx1}) and (\ref{constdx2}) in the Randall-Sundrum 
model for large $|c|$ 
since only the scalar fields in the hypermultiplet 
receive the mass shift which is linear in $w_\pi$.

\setcounter{equation}{0}
%%%%%%%%%%%%%%%%%%%%%%%%%%%%%%%%%%%%%%%%%%%%%%%%%%%%%%%%%
\section{Another approximation for background solution}
\label{sc:background}
%%%%%%%%%%%%%%%%%%%%%%%%%%%%%%%%%%%%%%%%%%%%%%%%%%%%%%%%%
In this appendix, we consider another approximation to 
explore solutions of the equations of motion.

We see that F components are nonlinear in $\phi, \phi^c$, 
resulting in a difficulty to solve the 
equations of motion for $\phi, \phi^c$. 
To examine the leading order effects of the 
constant boundary superpotential, we attempt 
the following approximation 
\bea
\phi, \phi^c &\sim& {\cal O}(R^{-3/2}), \qquad 
w_{0, \pi} \sim {\cal O}(1), 
\label{eq:apprx1}
\\
R^{-1} &<& M_5, \qquad 
R\sigma \sim {\cal O}(10\pi). 
\label{eq:apprx2}
\eea
Evaluating the order of magnitude of the terms 
in the equations of motion for auxiliary fields 
(\ref{Feom})-(\ref{FTeom}), the dominant terms are 
found to be 
\bea
F &\approx& -\frac{e^{-R\sigma}}{R}
\left[
-\partial_y \phi^{c\dag} 
+ \left( \frac{3}{2} + c \right)R \sigma' \phi^{c\dag}
+\frac{\phi}{2M_5^3}W_b \right], \label{Foemd}\\
F^c &\approx& -\frac{e^{-R\sigma}}{R}
\left[
\partial_y \phi^{\dag} 
- \left( \frac{3}{2} - c \right)R \sigma' \phi^{\dag}
+\frac{\phi^c}{2M_5^3}W_b \right], \label{Fceomd} \\
F_{\varphi} &\approx& -\frac{e^{-R\sigma}}{R}
\left[
-\frac{1}{6M_5^3}\phi^\dag \partial_y \phi^{c\dag} 
+\frac{1-6R\sigma}{6M_5^3}\phi^{c\dag}\partial_y \phi^\dag 
+\frac{3(1+R\sigma)
-2R \sigma c}{6M_5^3} \phi^\dag \phi^{c\dag} R\sigma'
-\frac{R\sigma}{M_5^3}W_b \right],
 \nonumber\\ 
\label{Fpeomd} \\
F_T &\approx& e^{-R\sigma}
\left[
\frac{1}{6M_5^3}\left\{ 6\phi^{c\dag} \partial_y \phi^\dag 
-2 \phi^{c\dag} \phi^\dag \left( \frac{3}{2}-c \right) R\sigma' 
+\frac{W_b}{M_5^3}(\phi^\dag \phi + \phi^{c\dag}\phi^c) \right\}
+\frac{W_b}{M_5^3} \right]. \label{FTeomd}
\eea
Then we find that the auxiliary fields Lagrangian is 
exactly given by the bilinear Lagrangian (\ref{auxap}).
Let us look for the classical solution for the background 
because the radion potential can be obtained from 
the classical solution. 
We assume the classical solution for the background 
to depend only on $y$. 
Since we obtain the bilinear Lagrangian (\ref{auxap}), 
the classical equations of motion is identical to the 
linearized one in (\ref{phieom}) and (\ref{phiceom}), 
except that $\partial_\mu\partial_\nu$ term in the 
last lines should be discarded. 

We can easily solve the equations of motion in the bulk 
\bea
\phi &=& e^{(3/2-c)R\sigma} [N_1 e^{(1+2c)R\sigma} + N_2], 
\label{cp}\\
h^c &=& e^{(3/2+c)R\sigma} [N_1^c e^{(1-2c)R\sigma} + N_2^c], 
\label{ch}
\eea
where $N_{1,2},N_{1,2}^c$ are integration constants. 
The boundary conditions from $\delta^2$ terms (\ref{BC1}) 
and (\ref{BC2}) 
become
\bea
\label{1}
0 &=& -2(N_1^c + N_2^c) +w_0 (N_1+N_2), \\
0 &=& 2e^{(3/2+c)Rk \pi} (N_1^c e^{(1-2c)Rk\pi} + N_2^c) 
+ w_\pi e^{(3/2-c)Rk\pi} (N_1 e^{(1+2c)Rk\pi} + N_2). 
\label{2}
\eea
The boundary conditions from $\delta$ terms (\ref{BC3}) 
and (\ref{BC4}) become
\bea
\label{3}
0 &=& -w_0 \left[ \frac{1}{3} 
\left( \frac{3}{2}+c \right) (N_1^c+N_2^c) 
+(1-2c) \frac{1}{3} N_1^c \right] \nonumber \\
&&+\frac{7}{3}w_0 (N_1^c+N_2^c) -2N_1 (1+2c), \\
0 
&=& -w_\pi \left[
\frac{1}{3} \left( \frac{3}{2} +c \right) 
(N_1^c e^{(1-2c)Rk\pi} + N_2^c)
+(1-2c)\frac{1}{3} N_1^c e^{(1-2c)Rk\pi} 
\right] \nonumber \\
&&+\frac{7}{3}(N_1^c e^{(1-2c)Rk\pi} 
+ N_2^c)+ 2N_1 (1+2c) e^{Rk\pi}. 
\label{4}
\eea
From (\ref{3}) and (\ref{4}), we obtain
\begin{eqnarray}
  \left(\begin{array}{c}
   N_1^c\\
   N_2^c
	\end{array}\right)
  &=& N_2 w_0 \left[
   -w_0^2 w_\pi {1-e^{(1-2c)Rk\pi}\over 12(1+2c)}
   +   
   {w_\pi+w_0 e^{Rk\pi}\over 9/2+c}
    -{w_\pi e^{(1-2c)Rk\pi}+w_0 e^{Rk\pi}\over
    11/2-c }
   \right]^{-1}
\nonumber
\\
 &&\times
   \left(
   \begin{array}{c}
  \left(w_\pi+w_0 e^{Rk\pi}\right) /(9+2c)
 \\
   -\left(w_\pi e^{(1-2c)Rk\pi}+w_0 e^{Rk\pi}\right)/ (11-2c)
     \end{array}\right). 
     \label{step1}
\end{eqnarray}

%\begin{eqnarray}
% \left(\begin{array}{c}
%       N_1^c\\
%       N_2^c\\
%    \end{array}\right)
%  &=&N_1{6(1+2c)\over 1-e^{(1-2c)Rk\pi}}
%   \left(\begin{array}{c}
%    (w_0^{-1}+w_\pi^{-1} e^{Rk\pi}) /(9/2+c)  \\
%    -(w_0^{-1}e^{(1-2c)Rk\pi}+w_\pi^{-1} e^{Rk\pi}) /(11/2-c)\\
%	 \end{array}\right). 
%  \label{step1}
%\end{eqnarray}
From (\ref{1}) and (\ref{step1}), we obtain
\begin{eqnarray}
N_1 &=& N_2 w_0 \left[
   -w_0^2 w_\pi {1-e^{(1-2c)Rk\pi}\over 12(1+2c)}
   +   
   {w_\pi+w_0 e^{Rk\pi}\over 9/2+c}
    -{w_\pi e^{(1-2c)Rk\pi}+w_0 e^{Rk\pi}\over
    11/2-c }
   \right]^{-1} \nonumber \\
&&\times w_0 w_\pi (1-e^{(1-2c)Rk\pi})/(12(1+2c)). 
\label{step2}
\end{eqnarray}

%\begin{eqnarray}
% N_2=N_1\left[
%  -1+{12(1+2c)w_0^{-1}\over 1-e^{(1-2c)Rk\pi}}
%       \left(
%  {w_0^{-1}+w_\pi^{-1}e^{Rk\pi}\over
%  9/2+c}-{w_0^{-1}e^{(1-2c)Rk\pi}+w_\pi^{-1}e^{Rk\pi}
%   \over 11/2-c}\right)\right]. 
%  \label{step2}
%\end{eqnarray}
It is interesting to see that 
$N_1^c,N_2^c,N_1\to 0$ for $w_0\to 0$.
We observe that the overall constant $N_2$ remains 
undetermined, because of the bilinear Lagrangian in our 
approximation. 
Moreover, we do not have good reasons to assume 
this constant $N_2$ not to depend on the 
radius $R$, since 
all these normalization constants 
$N_1, N_2, N^c_1, N^c_2$ are related by 
the boundary condition involving $R$. 
Therefore we cannot give a definite answer on the 
radius stabilization in the present approximation. 

If we take into account higher order corrections and/or 
additional mechanisms, 
%which are quite complicated, 
we can presumably determine the normalization constant,  
although it is a very hard task to analyze.

\end{appendix}

\vspace*{10mm}
%%%%%%%%%%%%% BIBLIOGRAPHY %%%%%%%%%%%%%%%%%%%%

%%%%%%%%%%%%%%%%%%%%%

\end{document}